\documentclass[aps,prd,groupaddress,floatfix,tighten,nofootinbib,showpacs,amsfonts,superscriptaddress]{revtex4}
\usepackage{url}
\usepackage{xcolor}
\usepackage{hyperref}

\colorlet{shadecolor}{gray!15}
\definecolor{greenLinks}{rgb}{0, 0.6, 0}
\definecolor{blueLinks}{rgb}{0, 0, 0.6}
\definecolor{redLinks}{rgb}{0.6, 0, 0}
\definecolor{tempText}{rgb}{0.55, 0.10,0.67}
\definecolor{eprintLinks}{rgb}{0.4, 0.4, 0.4}
\definecolor{journalLinks}{rgb}{0.6, 0, 0}
\usepackage{amssymb}
\usepackage{amsmath,color}
\usepackage{epsfig}
\usepackage{graphicx,epsf,epsfig}
\usepackage{footnote}
\usepackage{multirow}
\usepackage{enumitem}
\usepackage{color}
\def\slc#1{\setbox0=\hbox{$#1$}                  
    \dimen0=\wd0                                 
    \setbox1=\hbox{/} \dimen1=\wd1               
    \ifdim\dimen0>\dimen1                        
       \rlap{\hbox to \dimen0{\hfil/\hfil}}      
       #1                                        
    \else                                        
       \rlap{\hbox to \dimen1{\hfil$#1$\hfil}}   
       /                                         
    \fi}

\def\be{\begin{equation}}
\def\ee{\end{equation}}
\def\gs{\mathrel{
   \rlap{\raise 0.511ex \hbox{$>$}}{\lower 0.511ex \hbox{$\sim$}}}}
\def\ls{\mathrel{
   \rlap{\raise 0.511ex \hbox{$<$}}{\lower 0.511ex \hbox{$\sim$}}}}

\newcommand{\ba}{\begin{array}{c}}
\newcommand{\baz}{\begin{array}{cc}}
\newcommand{\barrr}{\begin{array}{rrr}}
\newcommand{\bad}{\begin{array}{ccc}}
\newcommand{\bav}{\begin{array}{cccc}}
\newcommand{\baf}{\begin{array}{ccccc}}
\newcommand{\bea}{\begin{equation} \begin{array}{c}}
\newcommand{\eea}{\end{array} \end{equation}}
\newcommand{\ea}{\end{array}}

\usepackage[normalem]{ulem}

\def\21{$\mathrm{SU(2)_L \otimes U(1)_Y}$ }

\newcommand {\ignore}[1]{}

\newcommand{\vt}{\vert}
\newcommand{\nn}{\nonumber}

\def\dg{\dagger}     

\allowdisplaybreaks \allowdisplaybreaks[2]
  
\newcommand{\AddrCECYT}{Centro de Estudios Cient\'ificos y Tecnol\'ogicos No 16, Instituto Polit\'ecnico Nacional, Pachuca: Ciudad del Conocimiento y la Cultura, Carretera Pachuca Actopan km 1+500, San Agust\'in Tlaxiaca, Hidalgo, M\'exico.\\ 
}

\begin{document}
\title{
A comparative study between the modified Fritzsch and nearest neighbor interaction textures} 
%
\author{J. D. Garc\'ia-Aguilar}
\email{jdgarcia@ipn.mx}
\affiliation{\AddrCECYT}

\author{Juan Carlos G\'omez-Izquierdo}
\email{jcgizquierdo1979@gmail.com}
\affiliation{\AddrCECYT}
%

%

%

\date{\bf \today} 

\begin{abstract}\vspace{2cm}
From mass textures point of view, we present a comparative study of the $\mathbf{S}_{3}$ flavor symmetry in the left-right symmetry model (LRSM) and the baryon minus lepton model (BLM) taking into account their predictions on the CKM mixing matrix. To do this, we recover the already studied quark mass matrix, that comes from some published papers, and under certain strong assumption, one can show that there are predictive scenarios in the LRSM and BLM where the modified Fritzsch and nearest neighbor interaction (NNI) textures drive respectively the quark mixings. As main result, the CKM mixing matrix is in good agreement with the last experimental data in the flavored BLM model.
\end{abstract}

%
\maketitle
%

\section{Introduction}
Understanding the peculiar patterns in the CKM~\cite{Cabibbo:1963yz, Kobayashi:1973fv} and PMNS~\cite{Maki:1962mu, Pontecorvo:1967fh} mixing matrices is still a challenge in the Standard Model (SM) and beyond it. As it is well known, in the quark sector, the CKM matrix is almost the identity one which differs completely to the PMNS matrix where large values, in its entries, can be found. 

The mass textures, in the fermion mass matrices, have been useful to try of explaining phenomenological the contrasting mixing matrices. In this line of though, the pronounced hierarchy among the quark masses,
$m_{t}\gg m_{c}\gg m_{u}$ and $m_{b}\gg m_{s}\gg m_{d}$, could be
behind the small mixing angles, that parametrize the CKM, which
depend strongly on the mass ratios \cite{Fritzsch:1999ee,Xing:2014sja, Verma:2015mgd}. This notable hierarchy may naturally come from the hierarchical mass matrices, as for example, the Fritzsch~ \cite{Fritzsch:1977za, Fritzsch:1977vd, Fritzsch:1979zq, Fritzsch:1985eg,Fritzsch:1999ee,Fritzsch:1999rb} and nearest neighbor interaction (NNI)~\cite{Branco:1988iq, Branco:1994jx, Harayama:1996am, Harayama:1996jr} mass textures. Although the former textures give us the extended Gatto-Sartori-Tonin relations
\cite{Gatto:1968ss,Cabibbo:1968vn,Oakes:1969vm,Fritzsch:1977za,Fritzsch:1977vd,Fritzsch:1979zq,Fritzsch:1999ee,Fritzsch:1999rb}, this framework
presents some problems with the top mass and the $V_{cb}$ element of
the CKM matrix, as can be
seen in~\cite{Branco:2010tx, Fritzsch:2011cu}. However, the NNI textures are capable of fitting with great accuracy the CKM mixing matrix. On the other hand, the lepton mixings may be reproduced quite well by the hierarchical mass matrices if the normal ordering were obeyed by the the neutrino masses, but, the inverted mass spectrum is not ruled out by the experiments so far. In short, this sector has to be treated with finesse.  

In the model building context, the mass textures have been generated dynamical by the flavor symmetries~\cite{Ishimori:2010au,Grimus:2011fk,Ishimori:2012zz,King:2013eh}. As a result of this, in the last years, a plethora of discrete symmetries have been proposed to accommodate the fermion mixings, all of this in different theoretical frameworks. In particular, the $\mathbf{S}_{3}$ flavor symmetry, and its implications on masses and mixings, has been studied in the left-right symmetry (LRSM)~\cite{Gomez-Izquierdo:2017rxi, Garces:2018nar} and baryon minus lepton number models (BLM)~\cite{Gomez-Izquierdo:2018jrx}.

In the current paper, from mass textures point of view, we present a comparative study of the $\mathbf{S}_{3}$ flavor symmetry in the left-right symmetry model (LRSM) and baryon minus lepton model (BLM) taking into account their predictions on the CKM mixing matrix. To do this, we will recover the already studied quark mass matrix, that comes from some published papers \cite{Gomez-Izquierdo:2017rxi, Garces:2018nar, Gomez-Izquierdo:2018jrx}, and under certain strong assumption, one can show that there are predictive scenarios in the LRSM and BLM where the modified Fritzsch and nearest neighbor interaction (NNI) textures drive respectively the quark mixings. As main result, the CKM mixing matrix is in good agreement with the last experimental data in the flavored BLM model.

The plan of the paper is as follows: two flavored gauge models are describe briefly in section II, along with this the quark mass matrix that comes from those models; in the section III, the corresponding quark mixing matrix are obtained and relevant features are remarked.
A numerical study is carried out, in section IV, to find a set of values the free parameters that accommodate the mixings. We close with relevant conclusions in section V.

%
%
%
%
%

\section{Quark Mass Matrix}

As it is well known, in several extended models with three Higgs doblets (3HD) and the $\mathbf{S}_{3}$ flavor symmetry \footnote{
The ${\bf S}_{3}$ flavor symmetry has been studied exhaustively in different frameworks~\cite{Pakvasa:1977in, Gerard:1982mm, Kubo:2003iw, Kubo:2003pd,Kobayashi:2003fh, Chen:2004rr, Kubo:2005sr, Felix:2006pn, Mondragon:2007af, Mondragon:2007nk, Mondragon:2007jx, Meloni:2010aw, Dicus:2010iq, Dong:2011vb, Canales:2011ug, Canales:2012ix, Kubo:2012ty, Canales:2012dr, GonzalezCanales:2012kj, Dias:2012bh, GonzalezCanales:2012za, Meloni:2012ci, Canales:2013ura, Ma:2013zca, Canales:2013cga, Hernandez:2014lpa, Hernandez:2014vta, Ma:2014qra, Gupta:2014nba, Hernandez:2015dga, Hernandez:2015zeh, Hernandez:2015hrt, Arbelaez:2016mhg, Hernandez:2013hea, CarcamoHernandez:2016pdu, Das:2014fea, Das:2015sca, Pramanick:2016mdp, Das:2017zrm, Cruz:2017add, Ge:2018ofp, Das:2018rdf, Xing:2019edp, Pramanick:2019oxb}. In most of these works, the meaning of the flavor has been extended to the scalar sector such that three Higgs doublets are required to fermion masses and mixings.
}, the quark mass matrix is given by
\begin{equation}
{\bf M}_{q}=\begin{pmatrix}
a_{q}+b^{\prime}_{q} & b_{q} & c_{q} \\ 
b_{q} & a_{q}-b^{\prime}_{q} & c^{\prime}_{q} \\ 
f_{q} & f^{\prime}_{q} & g_{q}
\end{pmatrix},\label{eq1} 
\end{equation}
where the $q= u, d$ and the matrix elements depend on the theoretical framework where the model is realized. In the current work, we will focus in two studied frameworks.

\subsection{Flavored left-right symmetric model (FLRSM)}
The minimal left-right symmetric model (LRSM) is based in the $SU(3)_{c}\otimes SU(2)_{L}\otimes SU(2)_{R}\otimes U(1)_{B-L}$ gauge group which is an appealing extension of the SM. The quark and scalar fields and their respective quantum numbers (in parenthesis) under the gauge symmetry are given by
\begin{eqnarray}
Q_{L}&=&\begin{pmatrix}
u\\ 
d
\end{pmatrix}_{L}\sim {\left(3, 2, 1, 1/3\right)},\qquad Q_{R}=\begin{pmatrix}
u\\ 
d
\end{pmatrix}_{R}\sim {\left(3, 1, 2, 1/3\right)},\qquad
\Phi=\begin{pmatrix}
\phi^{0} & \phi^{'+} \\
\phi^{-} & \phi^{'0} \\
\end{pmatrix} \sim \left(1, 2, 2, 0 \right).
\end{eqnarray}

Then, we have the the Yukawa mass term
\begin{equation}
-\mathcal{L}_{Y}=\bar{Q}_{L}\left[y^{q}\Phi+ \tilde{y}^{q}\tilde{\Phi} \right]Q_{R}+h.c. \label{yt}
\end{equation}
where the family indexes have been suppressed and $\tilde{\Phi}_{i}=-i\sigma_{2}\Phi^{\ast}_{i}i\sigma_{2}$. As it is well known, in the LRSM too many Yukawa ($y^{q}$ and $\tilde{y}^{q}$) couplings appear in the mass matrices, as can be seen in Eq. (\ref{yt}). This drawback may be alleviated by Parity Symmetry, $\Psi_{i L}\leftrightarrow \Psi_{i R}$ and
$\Phi_{i} \leftrightarrow \Phi^{\dg}_{i}$, which relates the Yukawa couplings ($y=y^{\dagger}$, $\tilde{y}=\tilde{y}^{\dagger}$) and the gauge couplings too. On the other hand, after the spontaneous symmetry breaking, $\langle\Phi\rangle=\textrm{Diag.}(k,~k^{\prime})$, the quark mass matrix are given as $\mathbf{M}_{u}=\mathbf{y}^{u}k+\tilde{\mathbf{y}}^{u}k^{\prime\ast}$ and $\mathbf{M}_{d}=\mathbf{y}^{d}k^{\prime}+\tilde{\mathbf{y}}^{d}k^{\ast}$. As a result of imposing Parity Symmetry, one will end up having a complex symmetric quark mass matrix if the vacuum expectation values (vev's) are complex; in the literature this scenario is well known as {\bf pseudomanifest left-right symmetry} (PLRT)~\cite{Branco:1982wp, Langacker:1989xa, Harari:1983gq}. If the vev's are real, the quark mass matrix is hermitian and the number of CP phases are reduced, this framework is known as {\bf manifest left-right symmetry} (MLRT)~\cite{Beg:1977ti, Langacker:1989xa}.

As was already commented, the LRSM framework was combined with the ${\bf S}_{3}$ flavor symmetry to provide a flavored non-minimal left-right symmetric model~\cite{Gomez-Izquierdo:2017rxi,Garces:2018nar}. In those papers, three Higgs bidoublets, $\Phi_{1, 2, 3}$, are required to accommodate the PMNS and CKM mixing matrices. Then, the matter content of the model transforms in a not trivial way under the ${\bf S}_{3}$ symmetry and this is displayed in the table below. Remarkably, the flavor assignation provides hierarchical mass matrices as we will see later.
\begin{table}[ht]
	\begin{center}
		\begin{tabular}{|c|c|c|c|c|}
			\hline\hline 
			{\footnotesize Matter}	& {\footnotesize $Q_{I (L, R)}$} & {\footnotesize $Q_{3 (L, R)}$} & {\footnotesize $\Phi_{I}$} &  {\footnotesize $\Phi_{3}$} \\ 
			\hline 
			{\footnotesize \bf $S_{3}$}	& {\footnotesize \bf $2$} & {\footnotesize \bf $1_{S}$}  & {\footnotesize \bf $2$} & {\footnotesize \bf $1_{S}$}  \\ 
			\hline\hline 
		\end{tabular} \caption{Flavored left-right symmetric model. Here, $I=1,2$.}
	\end{center}	
\end{table}

The lepton sector was leave aside in this particular scenario, so the full assignment can be found in~\cite{Gomez-Izquierdo:2017rxi, Garces:2018nar} for more details. In the mentioned papers, the  quark mass matrix, $\mathbf{M}_{q}$, has the entries
\begin{eqnarray}
a_{u}&=&y^{q}_{2}k_{3}+\tilde{y}^{q}_{2}k^{\prime \ast}_{3},\quad b^{\prime}_{u}=y^{q}_{1}k_{2}+\tilde{y}^{q}_{1}k^{\prime \ast}_{2},\quad b_{u}=y^{q}_{1}k_{1}+\tilde{y}^{q}_{1}k^{\prime \ast}_{1},\quad c_{u}=y^{q}_{3}k_{1}+\tilde{y}^{q}_{3}k^{\prime \ast}_{1};\nn\\
c^{\prime}_{u}&=&y^{q}_{3}k_{2}+\tilde{y}^{q}_{3}k^{\prime \ast}_{2},\quad f_{u}=y^{\dg q}_{3}k_{1}+\tilde{y}^{\dg q}_{3}k^{\prime \ast}_{1},\quad f^{\prime}_{u}=y^{\dg q}_{3}k_{2}+\tilde{y}^{\dg q}_{3}k^{\prime \ast}_{2},\quad g_{u}=y^{q}_{5}k_{3}+\tilde{y}^{q}_{5}k^{\prime \ast}_{3};\nn\\
a_{d}&=&y^{q}_{2}k^{\prime}_{3}+\tilde{y}^{q}_{2}k^{\ast}_{3},\quad b^{\prime}_{d}=y^{q}_{1}k^{\prime}_{2}+\tilde{y}^{q}_{1}k^{\ast}_{2},\quad b_{d}=y^{q}_{1}k^{\prime}_{1}+\tilde{y}^{q}_{1}k^{\ast}_{1},\quad c_{d}=y^{q}_{3}k^{\prime}_{1}+\tilde{y}^{q}_{3}k^{\ast}_{1};\nn\\
c^{\prime}_{d}&=&y^{q}_{3}k^{\prime}_{2}+\tilde{y}^{q}_{3}k^{\ast}_{2},\quad f_{d}= y^{\dg q}_{3}k^{\prime}_{1}+\tilde{y}^{\dg q}_{3}k^{\ast}_{1},\quad f^{\prime}_{d}= y^{\dg q}_{3}k^{\prime}_{2}+\tilde{y}^{\dg q}_{3}k^{\ast}_{2},\quad g_{d}=y^{q}_{5}k^{\prime}_{3}+\tilde{y}^{q}_{5}k^{\ast}_{3}.
\label{eq2}
\end{eqnarray}
where Parity Symmetry was imposed in the model so that both sceanrios, PLRT and MLRT, were studied in a particular alignment in the vev's, $k_{1}=k_{2}$ and $k^{\prime}_{1}=k^{\prime}_{2}$, to reduce the free parameters.

In the current work, we will recover the MLRT scenario, however, the study is totally different to that already presented in ~\cite{Gomez-Izquierdo:2017rxi, Garces:2018nar} since that a strong assumptions will be done in the vev's in order to reduce a little more the parameters. As a result of this, a simplest and predictive scenario comes out but some disadvantages will be appeared.

Henceforth, we assume that $k^{
\prime}_{i}=0$, then the quark mass matrix given in Eq. (\ref{eq2}) contains fewer Yukawa couplings and that will be denoted by
\begin{equation}
{\bf \mathcal{M}}_{q}\approx\begin{pmatrix}
a_{q}+b^{\prime}_{q} & b_{q} & c_{q} \\ 
b_{q} & a_{q}-b^{\prime}_{q} & c^{\prime}_{q} \\ 
c^{\ast
}_{q} & c^{\prime\ast}_{q} & g_{q}
\end{pmatrix}.\label{eq3} 
\end{equation}

In what follows, we will show how to above mass matrix possesses implicitly a kind of Fritzsch textures but an extra free parameter, in two entries of the diagonal, modify that textures.

 \subsection{Flavored baryon minus lepton model (FBLM)}

An other interesting extension of the SM is the well known B-L gauge model, this is based on the $SU(3)_{c}\otimes SU(2)_{L}\otimes
U(1)_{Y}\otimes U(1)_{B-L}$ gauge group where, apart from the SM
fields, three $N_{i}$ RHN's and a $\phi$ singlet scalar field are
added to the matter content. Additionally, this framework provides other nice features that SM does not has it, as for example, the RHN's get their mass \textit{a la Higgs} and the type I see-saw mechanism arises in naturally way.

Under B-L, the quantum numbers for
quarks, leptons and Higgs ($\phi$) are $1/3$, $-1$ and $0$ ($-2$),
respectively. Then, the gauge invariant Lagrangian is

\begin{equation}
\mathcal{L}_{B-L}=\mathcal{L}_{SM}-y^{D}\bar{L}\tilde{H}N-\frac{1}{2}y^{N}\bar{N}^{c}\phi N.
\end{equation}

Within this theoretical framework, the $\mathbf{S}_{3}$ flavor symmetry was also explored and the three Higgs doublets were included to generate the fermion mixings. In particular, in the model \cite{Gomez-Izquierdo:2018jrx}, the quark and Higgs fields transform in a not trivial form as can be seen in the following table. 

\begin{table}[ht]
\begin{center}
\begin{tabular}{|c|c|c|c|c|c|c|c|c|c|c|c|c|c|c|}
\hline \hline	
			{\footnotesize Matter} & {\footnotesize $Q_{I L}, H_{I}, d_{I R}, u_{I R}$ } & {\footnotesize $Q_{3 L}, H_{3}, d_{3 R}, u_{3 R}$}  \\ \hline
			{\footnotesize \bf $S_{3}$} &  {\footnotesize \bf $2$} & {\footnotesize \bf $1_{S}$}  \\ \hline \hline
		\end{tabular}\caption{Flavored B-L model. Here $I=1,2$.}\label{tab2}
	\end{center}
\end{table}

Given the above assignation, the quark mass matrix has the same form as that given in Eq. (\ref{eq1}). In this scenario, the quark matrix will be denoted by

\begin{equation}
\bar{{\bf \mathcal{M}}}_{q}=\begin{pmatrix}
a_{q}+b^{\prime}_{q} & b_{q} & c_{q} \\ 
b_{q} & a_{q}-b^{\prime}_{q} & c^{\prime}_{q} \\ 
f_{q} & f^{\prime}_{q} & g_{q}
\end{pmatrix}.\label{eqBL} 
\end{equation}

where the matrix elements are
\begin{eqnarray}
a_{q}&=&y^{q}_{2}\langle H_{3}\rangle,\quad b^{\prime}_{q}=y^{q}_{1} \langle H_{2}\rangle, \quad b_{q}=y^{q}_{1} \langle H_{1}\rangle,\quad c_{q}=y^{q}_{3} \langle H_{1}\rangle;\nn\\
c^{\prime}_{q}&=&y^{q}_{3} \langle H_{2}\rangle,\quad f_{q}=y^{q}_{4} \langle H_{1}\rangle;\quad f^{\prime}_{q}= y^{q}_{4} \langle H_{2}\rangle 
,\quad g_{q}=y^{q}_{5} \langle H_{3}\rangle,\quad
\label{eq9}
\end{eqnarray}
where the $q=u, d$. In here, it is convenient to remark the number of Yukawa couplings is reduced to half in comparison to the FLRSM~\cite{Gomez-Izquierdo:2017rxi, Garces:2018nar}. Along with this, the $\bar{{\bf \mathcal{M}}}_{q}$ quark mass, in general, is not hermitian neither complex symmetric so that extra free parameters must contain the quark mass matrix. As we will see, this possesses implicitly the NNI textures that are capable to accommodate the CKM mixings.

\section{Quark Mixing Matrix}

In this section, we will show explicitly that the quark mass matrix possesses implicitly the Fritzsch and the NNI textures in the FLRSM and FBLM, respectively. To do this, we have to keep in mind that ${\bf \mathcal{M}}_{q}$ and $\bar{{\bf \mathcal{M}}}_{q}$ are diagonalized as $\mathcal{U}^{\dagger}_{q L} {\bf \mathcal{M}}_{q} \mathcal{U}_{q L}=\hat{\mathbf{M}}_{q}$ and ${\bf U}^{\dagger}_{q L} \bar{{\bf \mathcal{M}}}_{q}
{\bf U}_{q R}=\hat{\bf M}_{q}$ where $\mathbf{\hat{M}}_{q}=\textrm{Diag.}(m_{q_{1}}, m_{q_{2}}, m_{q_{3}})$ with $m_{q_{i}}$ being the physical quark masses. Then, we make the following rotation for each case $\mathcal{U}_{q L}=\mathbf{U}_{\theta} \mathbf{u}_{q }$ and ${\bf U}_{q(L, R)}={\bf U}_{\theta} {\bf u}_{q (L, R)}$
so that one obtains $\mathbf{u}^{\dagger}_{q} \mathbf{ m}_{q}\mathbf{u}_{q}=\mathbf{\hat{M}}_{q}$ and $\mathbf{u}^{\dagger}_{q L} \bar{\mathbf{m}}_{q}\mathbf{u}_{q R}=\mathbf{\hat{M}}_{q}$ where $\mathbf{m}_{q}$, $\bar{\mathbf{m}}_{q}$ and $\mathbf{U}_{\theta}$ are given respectively as

\begin{equation}
\mathbf{m}_{q}=\begin{pmatrix}
a_{q} & \frac{2}{\sqrt{3}}b^{\prime}_{q} & 0 \\ 
\frac{2}{\sqrt{3}}b^{\prime}_{q} & a_{q} & \frac{2}{\sqrt{3}}c^{\prime}_{q} \\ 
0 & \frac{2}{\sqrt{3}}c^{\prime\ast}_{q} & g_{q}
\end{pmatrix},\qquad 
\bar{{\bf m}}_{q}=
\begin{pmatrix}
a_{q} & \frac{2}{\sqrt{3}}b^{\prime}_{q} & 0 \\ 
\frac{2}{\sqrt{3}}b^{\prime}_{q} & a_{q} & \frac{2}{\sqrt{3}}c^{\prime}_{q} \\ 
0 & \frac{2}{\sqrt{3}}f^{\prime}_{q} & g_{q}
\end{pmatrix},
\qquad\mathbf{U}_{\theta}= \begin{pmatrix}
\cos{\theta} & \sin{\theta} & 0 \\ 
-\sin{\theta} & \cos{\theta} & 0 \\ 
0 & 0 & 1
\end{pmatrix}\label{eq4} 
\end{equation}

with the following two conditions to be satisfied

\begin{eqnarray}
\tan{\theta}&=&\frac{c_{q}}{c^{\prime}_{q}}=\frac{c^{\ast}_{q}}{c^{\prime\ast}_{q}}=\frac{k_{1}}{k_{2}}\qquad \textrm{and}\qquad \tan{2\theta}=\frac{b^{\prime}_{q}}{b_{q}}=\frac{k_{2}}{k_{1}}\qquad \textrm{FLRSM}\nonumber\\
\tan{\theta}&=&\frac{c_{q}}{c^{\prime}_{q}}=\frac{f_{q}}{f^{\prime}_{q}}=\frac{\langle H_{1}\rangle}{\langle H_{2}\rangle}\qquad \textrm{and}\qquad \tan{2\theta}=\frac{b^{\prime}_{q}}{b_{q}}=\frac{\langle H_{2}\rangle}{\langle H_{1}\rangle} \qquad \textrm{FBLM}
\label{eq5}
\end{eqnarray}

Both relations give us the relation $k_{2}=\pm\sqrt{3}k_{1}$ ($\langle H_{2}\rangle=\pm\sqrt{3} \langle H_{1}\rangle$), therefore $\theta=\pi/6$ or $5\pi/6$ as was shown in
\cite{Canales:2013cga}. As can be noticed, for each case, the diagonalization proceed will be different due to the nature of the quark mass matrix.

\subsection{FLRSM}
The $\mathbf{m}_{q}$ mass matrix (given in Eq. (\ref{eq4})) may be written as

\begin{equation}
\mathbf{m}_{q}=\begin{pmatrix}
a_{q} & B_{q} & 0 \\ 
B_{q} & a_{q} & C_{q} \\ 
0 & C^{\ast}_{q} & g_{q}
\end{pmatrix} \label{eq6} 
\end{equation}
where $B_{q}=2b^{\prime}_{q}/\sqrt{3}$ and $C_{q}=2c^{\prime}_{q}/\sqrt{3}$. Without loss of generality, we assume that $a_{q}$, $B_{q}$ and $g_{q}$ are positive defined. Then,
as one can realize, if $a_{q}$ was zero, the mass matrix would possess the Fritzsch textures with one CP violation phase less since that $B_{q}$ is real. Then, this extra parameter, $a_{q}$, will modify slightly the Fritzsch textures as we will see in the diagonalization proceed. In order to do this, for simplicity, the matrix elements in the Eq. (\ref{eq6}) will be normalized by the heaviest mass. 

Having commented the CP violating phases can be factorized as $\mathcal{P}_{q}\tilde{\mathbf{m}}_{q}\mathcal{P}^{\dagger}_{q}$ where $\mathcal{P}_{q}=\textrm{Diag}.(1, 1,e^{-i\alpha_{C_{q}}} )$ with $\alpha_{C_{q}}=arg(C_{q})$ and $\tilde{\mathbf{m}}_{q}$ being a real symmetric matrix. Thus, we choose appropriately $\mathbf{u}_{q}=\mathcal{P}_{q}\mathcal{O}_{q}$ where the latter real orthogonal matrix is given as

\begin{equation}
\mathcal{O}_{q}=\begin{pmatrix}
\sqrt{\frac{( \tilde{m}_{q_{2}}+\tilde{a}_{q})(1-\tilde{a}_{q})\mathcal{M}_{q_{2}}}{\mathcal{D}_{q_{1}}}}& -\sqrt{\frac{(\tilde{m}_{q_{1}}-\tilde{a}_{q})(1-\tilde{a}_{q})
		\mathcal{M}_{q_{1}}}{\mathcal{D}_{q_{2}}}}
& \sqrt{\frac{(\tilde{m}_{q_{2}}+\tilde{a}_{q})(\tilde{m}_{q_{1}}-\tilde{a}_{q})\mathcal{M}_{q_{3}}}{\mathcal{D}_{q_{3}}}} 
\\ 
\sqrt{\frac{(\tilde{m}_{q_{1}}-\tilde{a}_{q})\mathcal{M}_{q_{2}}\mathcal{D}_{q}}{\mathcal{D}_{q_{1}}}}& \sqrt{\frac{(\tilde{m}_{q_{2}}+\tilde{a}_{q})\mathcal{M}_{q_{1}}\mathcal{D}_{q}}{\mathcal{D}_{q_{2}}}}
& \sqrt{\frac{(1-\tilde{a}_{q})\mathcal{M}_{q_{3}}\mathcal{D}_{q}}{\mathcal{D}_{q_{3}}}} \\ 
-\sqrt{\frac{(\tilde{m}_{q_{1}}-\tilde{a}_{q})\mathcal{M}_{q_{1}}\mathcal{M}_{q_{3}}}{\mathcal{D}_{q_{1}}}}&-\sqrt{\frac{(\tilde{m}_{q_{2}}+\tilde{a}_{q})\mathcal{M}_{q_{2}}\mathcal{M}_{q_{3}}}{\mathcal{D}_{q_{2}}}} 
& 
\sqrt{\frac{(1-\tilde{a}_{q})\mathcal{M}_{q_{1}}\mathcal{M}_{q_{2}}}{\mathcal{D}_{q_{3}}}}
\label{eq7}
\end{pmatrix} 
\end{equation}
with 
\begin{eqnarray}
\mathcal{M}_{q_{1}}&=&1+\tilde{m}_{q_{1}}-2\tilde{a}_{q},\quad 
\mathcal{M}_{q_{2}}=1-\tilde{m}_{q_{2}}-2\tilde{a}_{q},\quad
\mathcal{M}_{q_{3}}=\tilde{m}_{q_{2}}-\tilde{m}_{q_{1}}+2\tilde{a}_{q}
,\quad
\mathcal{D}_{q}=1-\tilde{m}_{q_{2}}+\tilde{m}_{q_{1}}-3\tilde{a}_{q};\nonumber\\
\mathcal{D}_{q_{1}}&=&(1-\tilde{m}_{q_{1}})( \tilde{m}_{q_{2}}+\tilde{m}_{q_{1}})\mathcal{D}_{q},\quad
\mathcal{D}_{q_{2}}=(1+\tilde{m}_{q_{2}})( \tilde{m}_{q_{2}}+\tilde{m}_{q_{1}})\mathcal{D}_{q},\quad
\mathcal{D}_{q_{3}}=(1+ \tilde{m}_{q_{2}})(1-\tilde{m}_{q_{1}})\mathcal{D}_{q},
\label{eq8}
\end{eqnarray}
where $\tilde{m}_{q_{2}}=\vert m_{q_{2}}\vert /m_{q_{3}}$, $\tilde{m}_{q_{1}}=m_{q_{1}}/m_{q_{3}}$ and $\tilde{a}_{q}=a_{q}/ m_{q_{3}}$. In this parametrization, there is a constraint among the quark masses and the free parameter $\tilde{a}_{q}$, this is $1>\tilde{m}_{q_{2}}>\tilde{m}_{q_{1}}>\tilde{a}_{q}>0$. Finally, the mass matrices, that take place in the CKM one, are give as
$\mathcal{U}_{q L}= \mathbf{U}_{\theta}\mathcal{P}_{q}\mathcal{O}_{q}$. Therefore,

\begin{equation}
\mathbf{V}_{CKM}=\mathcal{U}^{\dagger}_{u} \mathcal{U}_{d}= \mathcal{O}^{T}_{u} \bar{\mathcal{P}}_{q}\mathcal{O}_{d},\quad \textrm{with}\quad \mathcal{\bar{P}}_{q}\equiv \mathcal{P}^{\dagger}_{u}\mathcal{P}_{d}=\textrm{Diag}. (1, 1, e^{-i\bar{\alpha}_{q} })\label{eq12}
\end{equation}

Remarkable comments have to do for pointing out the salient features of this simplest scenario (MLRT). Under this approach, the CKM mixing matrix has three free parameters $\tilde{a}_{u}$, $\tilde{a}_{d}$ and the CP violating phase, $\bar{\alpha}_{q}=\alpha_{C_{d}}-\alpha_{C_{u}}$. As a result, this scenario turns out being quite predictive.

In order to show that the extended Gatto-Sartori-Tonin relations are obtained in this framework, let us make an analytic study on the CKM mixing matrix. To do this, we will take particular values for the free parameter $\tilde{a}_{q}$ that satisfies the constraint $1>\tilde{m}_{q_{2}}>\tilde{m}_{q_{1}}>\tilde{a}_{q}>0$:
\begin{enumerate}
	\item If $\tilde{a}_{q}$ goes to $\tilde{m}_{q_{1}}$, we obtain
	\begin{equation}
	\mathcal{O}_{q}\approx
	\begin{pmatrix}
	1 & 0 & 0 \\ 
	0 & \sqrt{\frac{1-\tilde{m}_{q_{1}}}{1+\tilde{m}_{q_{2}}}} & \sqrt{\frac{\tilde{m}_{q_{2}}+\tilde{m}_{q_{1}}}{1+\tilde{m}_{q_{2}}}} \\ 
	0 &-\sqrt{\frac{\tilde{m}_{q_{2}}+\tilde{m}_{q_{1}}}{1+\tilde{m}_{q_{2}}}} & \sqrt{\frac{1-\tilde{m}_{q_{1}}}{1+\tilde{m}_{q_{2}}}} 
	\end{pmatrix}
	\end{equation}
	
	In this limit, the CKM mixing matrix would not be accommodated quite well.
	
	\item If $\tilde{a}_{q}$ goes to $0$, then
	as one can verify straight, one obtains the Fritzsch textures with
	
	\begin{equation}
	\mathbf{O}_{q}\approx
	\begin{pmatrix}
	\sqrt{\frac{\tilde{m}_{q_{2}}\left(1-\tilde{m}_{q_{2}}\right)}{
			\mathcal{D}_{q_{1}}}} & -\sqrt{\frac{\tilde{m}_{q_{1}} \left(1+\tilde{m}_{q_{1}}\right) }{\mathcal{D}_{q_{2}}}}& \sqrt{\frac{\tilde{m}_{q_{1}}\tilde{m}_{q_{2}} \left(\tilde{m}_{q_{2}}-\tilde{m}_{q_{1}}\right) }{\mathcal{D}_{q_{3}}}}  \\ 
	\sqrt{\frac{\tilde{m}_{q_{1}}\left(1-\tilde{m}_{q_{2}}\right) \left(1-\tilde{m}_{q_{2}}+\tilde{m}_{q_{1}}\right) }{\mathcal{D}_{q_{1}}}}  & \sqrt{\frac{\tilde{m}_{q_{2}}\left(1+\tilde{m}_{q_{1}}\right) \left(1-\tilde{m}_{q_{2}}+\tilde{m}_{q_{1}}\right) }{\mathcal{D}_{q_{2}}}} & \sqrt{\frac{\left(\tilde{m}_{q_{2}}-\tilde{m}_{q_{1}}\right) \left(1-\tilde{m}_{q_{2}}+\tilde{m}_{q_{1}}\right) }{\mathcal{D}_{q_{3}}}}\\ 
	-\sqrt{\frac{\tilde{m}_{q_{1}}\left(\tilde{m}_{q_{2}}-\tilde{m}_{q_{1}}\right) \left(1+\tilde{m}_{q_{1}}\right) }{\mathcal{D}_{q_{1}}}}  & -\sqrt{\frac{\tilde{m}_{q_{2}}\left(\tilde{m}_{q_{2}}-\tilde{m}_{q_{1}}\right) \left(1-\tilde{m}_{q_{2}}\right) }{\mathcal{D}_{q_{2}}}}  & \sqrt{\frac{\left(1-\tilde{m}_{q_{2}}\right) \left(1+\tilde{m}_{q_{1}}\right) }{\mathcal{D}_{q_{3}}}}
	\end{pmatrix}
	\end{equation}
If both sectors, $q=u, d$, have the same limit, one obtains the following relations
	\begin{eqnarray}
	V_{us} &\approx&-\sqrt{\frac{\tilde{m}_{d}}{\tilde{m}_{s} }}+\sqrt{\frac{\tilde{m}_{u}}{\tilde{m}_{c}}};\nn\\
	V_{ub}&\approx& \tilde{m}_{s} \sqrt{ \tilde{m}_{d}}+\sqrt{\tilde{m}_{s}}\sqrt{\frac{\tilde{m}_{u}}{ \tilde{m}_{c}}}-\sqrt{\tilde{m}_{u}}~e^{-i\bar{\alpha}_{q}};\nonumber\\
	V_{td}&\approx&\tilde{m}_{c} \sqrt{\tilde{m}_{u}}-\sqrt{\tilde{m}_{d}}~e^{-i\bar{\alpha}_{q}}+\sqrt{\tilde{m}_{c} }\sqrt{\frac{\tilde{m}_{d}}{\tilde{m}_{s} }};\nn\\
	V_{cb} &\approx& \sqrt{\tilde{m}_{s} }-\sqrt{\tilde{m}_{c}}~e^{-i\bar{\alpha}_{q}},
	\end{eqnarray}
which look like the Gatto-Sartori-Tonin relations. Additionally, one does notice that those entries do not depend strongly on the CP violating phase, $\bar{\alpha}_{q}$. This fact has a direct consequence on the Jarlskog invariant, as we will see in the numerical analysis.

\end{enumerate}

\subsection{FBLM}

In this case, see Eq.(\ref{eq4}), the quark mass matrix, $\bar{\mathbf{m}}_{q}$, contains five free parameters 

\begin{equation}
\bar{\mathbf{m}}_{q}=\begin{pmatrix}
a_{q} & B_{q} & 0 \\ 
B_{q} & a_{q} & C_{q} \\ 
0 & F_{q} & g_{q}
\end{pmatrix} \label{eq6} 
\end{equation}
which only three can be fixed in terms ($\vert B_{q}\vert$, $\vert C_{q}\vert$  and $\vert F_{q}\vert$) of the physical masses. Therefore, this scenario is not very predictive, however, for simplicity, let us adopt
the benchmark where $a_{q}=0$ which means that $y^{q}_{2}=0$. In this
way, the NNI textures arises in the quark mass matrix.

In this benchmark, we have to figure out the form of the ${\bf u}_{f R}$ and ${\bf u}_{f L}$
unitary matrices that diagonalize $\bar{{\bf m}}_{q}$. Then, we must build the
bilineal forms: ${\bf \hat{M}}_{q} {\bf \hat{M}}^{\dagger}_{q}={\bf
	u}^{\dagger}_{q L} \bar{{\bf m}}_{q} \bar{{\bf m }}^{\dagger}_{q} {\bf u}_{q L}$
and ${\bf \hat{M}}^{\dagger}_{q} {\bf \hat{M}}_{q}={\bf
	u}^{\dagger}_{q R} \bar{{\bf m}}^{\dagger}_{q} \bar{{\bf m }}_{q} {\bf u}_{q
	R}$, however, in this work we will only need to  obtain the
${\bf u}_{q L}$ left-handed matrix which takes place in the CKM
matrix. This is given by ${\bf u}_{q L}={\bf Q}_{q L}{\bf O}_{q L}$
where the former matrix contains the CP-violating phases, $ {\bf
	Q}_{q} = \textrm{diag} \left( 1,\exp i\alpha_{q}, \exp
i\beta_{q} \right)$, that comes from $\bar{{\bf m}}_{q} \bar{{\bf m
}}^{\dagger}_{q}$. ${\bf O}_{q L}$ is a real orthogonal matrix and this
is parametrized as
\begin{align} \label{ortho}
{\bf O}_{q L}= 
\begin{pmatrix}
-\sqrt{\dfrac{\tilde{m}_{q_{2}} (\rho^{q}_{-}-R^{q}) K^{q}_{+}}{4 y_{q} \delta^{q}_{1} \kappa^{q}_{1} }} & -\sqrt{\dfrac{\tilde{m}_{q_{1}} (\sigma^{q}_{+}-R^{q}) K^{f}_{+}}{4 y_{q} \delta^{q}_{2} \kappa^{q}_{2} }}  & \sqrt{\dfrac{\tilde{m}_{q_{1}} \tilde{m}_{q_{2}} (\sigma^{q}_{-}+R^{q}) K^{q}_{+}}{4 y_{q} \delta^{q}_{3} \kappa^{q}_{3} }} \\ 
-\sqrt{\dfrac{\tilde{m}_{q_{1}} \kappa^{q}_{1} K^{q}_{-}}{\delta^{q}_{1}(\rho^{q}_{-}-R^{q}) }} & \sqrt{\dfrac{\tilde{m}_{q_{2}} \kappa^{q}_{2} K^{q}_{-}}{\delta^{q}_{2}(\sigma^{q}_{+}-R^{q}) }} & \sqrt{\dfrac{\kappa^{q}_{3} K^{q}_{-}}{\delta^{q}_{3}(\sigma^{q}_{-}+R^{q}) }} \\ 
\sqrt{\dfrac{\tilde{m}_{q_{1}} \kappa^{q}_{1}(\rho^{q}_{-}-R^{q})}{2 y_{q}\delta^{q}_{1}}} & -\sqrt{\dfrac{\tilde{m}_{q_{2}} \kappa^{q}_{2}(\sigma^{q}_{+}-R^{q})}{2 y_{q}\delta^{q}_{2}}}  & \sqrt{\dfrac{\kappa^{q}_{3}(\sigma^{q}_{-}+R^{q})}{2 y_{q}\delta^{q}_{3}}}
\end{pmatrix}
\end{align}
with
\begin{align} \label{defortho}
\rho^{q}_{\pm}&\equiv 1+\tilde{m}^{2}_{q_{2}}\pm\tilde{m}^{2}_{q_{1}}-y^{2}_{q},\quad \sigma^{q}_{\pm}\equiv 1-\tilde{m}^{2}_{q_{2}}\pm(\tilde{m}^{2}_{q_{1}}-y^{2}_{q}),\quad
\delta^{q}_{(1, 2)}\equiv (1-\tilde{m}^{2}_{q_{(1, 2)}})(\tilde{m}^{2}_{q_{2}}-\tilde{m}^{2}_{q_{1}});\nn\\
\delta^{q}_{3}&\equiv (1-\tilde{m}^{2}_{q_{1}})(1-\tilde{m}^{2}_{q_{2}}),\quad \kappa^{q}_{1} \equiv  \tilde{m}_{q_{2}}-\tilde{m}_{q_{1}}y_{q},\quad \kappa^{q}_{2}\equiv \tilde{m}_{q_{2}}y_{q}-\tilde{m}_{q_{1}},\quad \kappa^{q}_{3}\equiv y_{q}-\tilde{m}_{q_{1}}\tilde{m}_{q_{2}};\nn\\
R^{q}&\equiv \sqrt{\rho^{q 2}_{+}-4(\tilde{m}^{2}_{q_{2}}+\tilde{m}^{2}_{q_{1}}+\tilde{m}^{2}_{q_{2}}\tilde{m}^{2}_{q_{1}}-2\tilde{m}_{q_{1}}\tilde{m}_{q_{2}}y_{q})},\quad	K^{q}_{\pm} \equiv  y_{q}(\rho^{q}_{+}\pm R^{q})-2\tilde{m}_{q_{1}}\tilde{m}_{q_{2}}.
\end{align}

In the above expressions, all the parameters have been normalized by
the heaviest physical quark mass, $m_{q_{3}}$. Along with this, from
the above parametrization, $y_{q}\equiv \vt g_{q}\vt/m_{q_{3}}$, is
the only dimensionless free parameter that cannot be fixed in terms of
the physical masses, but is is constrained by, 
$1>y_{q}>\tilde{m}_{q_{2}}>\tilde{m}_{q_{1}}$. Therefore, the
left-handed mixing matrix that takes places in the CKM matrix is given
by ${\bf U}_{q L}= {\bf U}_{\theta}{\bf Q}_{q}{\bf O}_{q L}$ where
$q=u, d$. Finally, the CKM mixing matrix is written as
\begin{equation}
{\bf V}_{PMNS}={\bf O}^{T}_{u L}{\bf P}_{q} {\bf O}_{d L}, \quad {\bf P}_{q}={\bf Q}^{\dagger}_{u}{\bf Q}_{d}=\textrm{diag.}\left(1, e^{i\alpha_{q}}, e^{i\beta_{q}} \right).
\end{equation}

This CKM mixing matrix has four free parameters, namely $y_{u}$,
$y_{d}$, and two phases $\alpha_{q}$ and $\beta_{q}$; these parameters will be constrained numerically in the next section. We ought to comment that, in this case, the analytic study will not be realized since that NNI textures have been studied quite well and those work perfectly in the quark sector~\cite{Branco:1988iq, Branco:1994jx, Harayama:1996am, Harayama:1996jr}.

\section{Numerical results}

In the following numerical analysis, for both scenarios, we will make some scattering plots to constraint the allowed region for the free parameters. To do this, we demand that the free parameters fit the CKM matrix elements $\vert V_{ud}\vert=0.97446\pm 0.00010$,  $\vert V_{us}\vert=0.97446\pm 0.00010$,  $\vert V_{ub}\vert=0.00365\pm 0.00012$ and $\vert V_{cb}\vert=0.04214\pm 0.00076$ up to $3~\sigma$ of their experimental values. Along with this, the physical quark masses will be considered as inputs at the top mass quark scale:
$\tilde{m}_{u} =(1.33 \pm 0.73)\times 10^{-5}$, $\tilde{m}_{c}=(3.91 \pm 0.42)\times 10^{-3}$, $\tilde{m}_{d}=(1.49 \pm 0.39)\times 10^{-3}$ and
$\tilde{m}_{s}=(2.19 \pm 0.53)\times 10^{-2}$. Having made that, we will see the predictions on the rest of the CKM entries and the Jarlskog invariant which is defined as $\mathcal{J}= \mathcal{I}m\left[V_{us}V_{cb}V^{\ast
}_{cs}V^{\ast}_{ub}\right]$. In the standard parametrization of the CKM mixing matrix, we have
\begin{equation}
\mathcal{I}m\left[V_{us}V_{cb}V^{\ast
}_{cs}V^{\ast}_{ub}\right]=\cos\theta_{12} \cos\theta_{23} \cos^{2}\theta_{13} \sin\theta_{12} \sin\theta_{23} \sin\theta_{13}\sin\delta.
\end{equation}
where $\theta_{12}$, $\theta_{13}$ and $\theta_{23}$ stand for the three mixing angles and $\delta$ the only CP violating phase.

\subsection{FLRSM}
In this simplified scenario, there are only three free parameters namely $\tilde{a}_{u}$, $\tilde{a}_{d}$ and the CP violating phase $\alpha_{q}$. These have to satisfy the following constraints $1>\tilde{m}_{q_{2}}>\tilde{m}_{q_{1}}>\tilde{a}_{q}>0$ with $q=u, d$ and $2\pi\geq\bar{\alpha}_{q}\geq 0$. With these conditions, we obtained the following scattering plots. 

\begin{figure}[h!]
\centering
\includegraphics[scale=0.465]{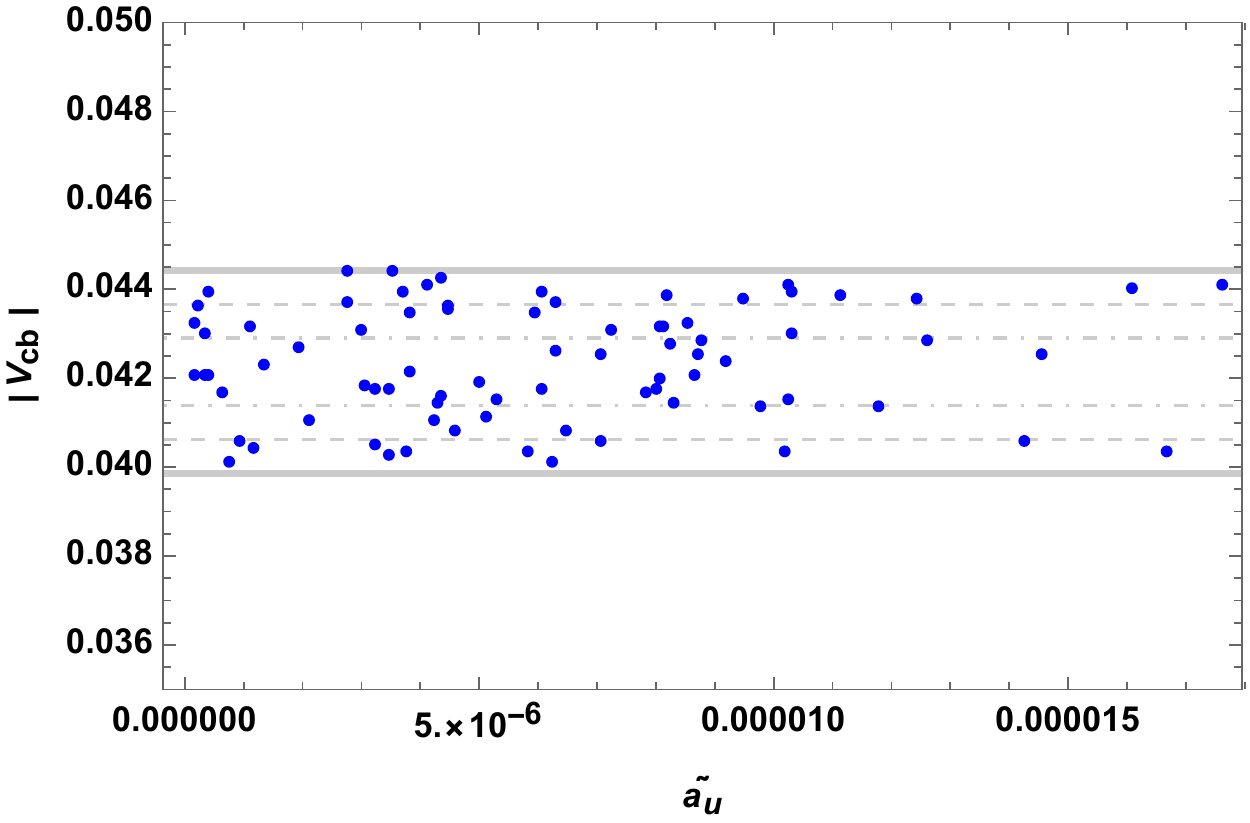}\hspace{1mm}\includegraphics[scale=0.465]{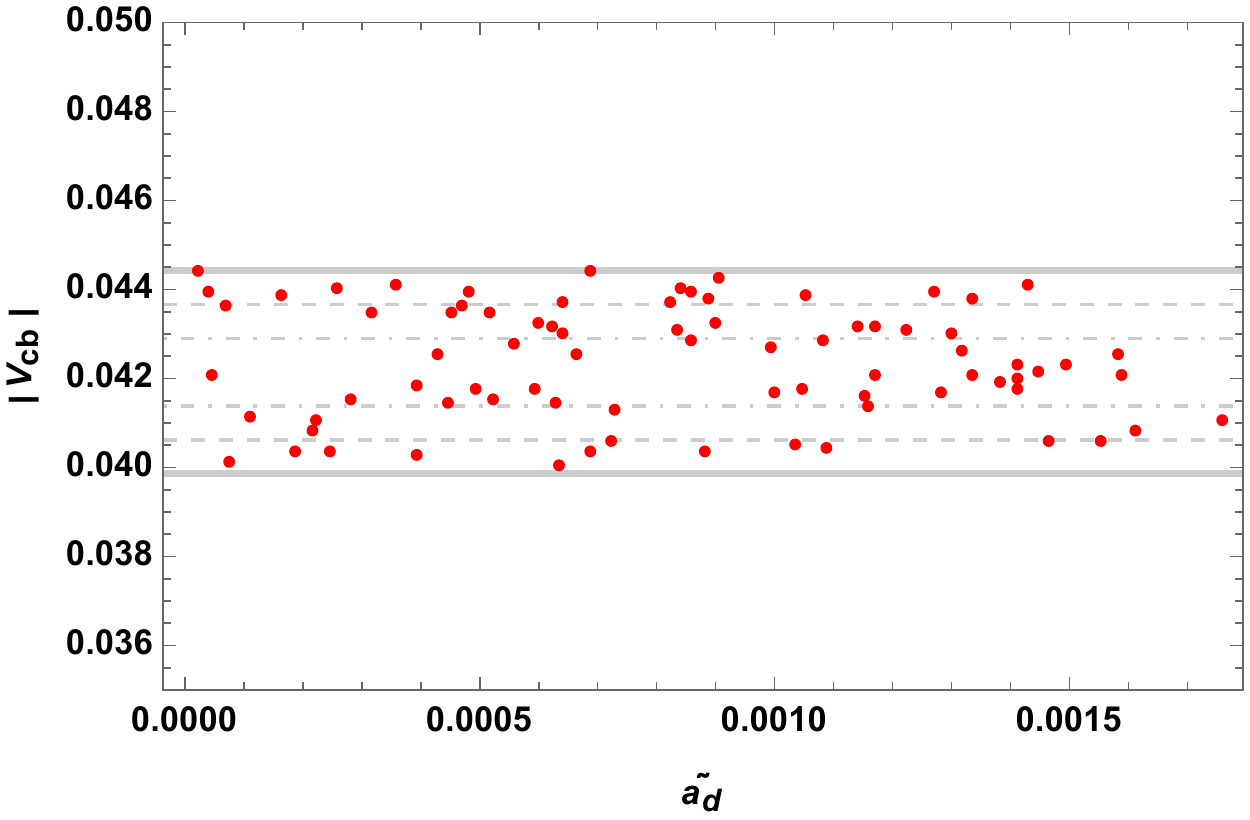}\hspace{1mm}\includegraphics[scale=0.465]{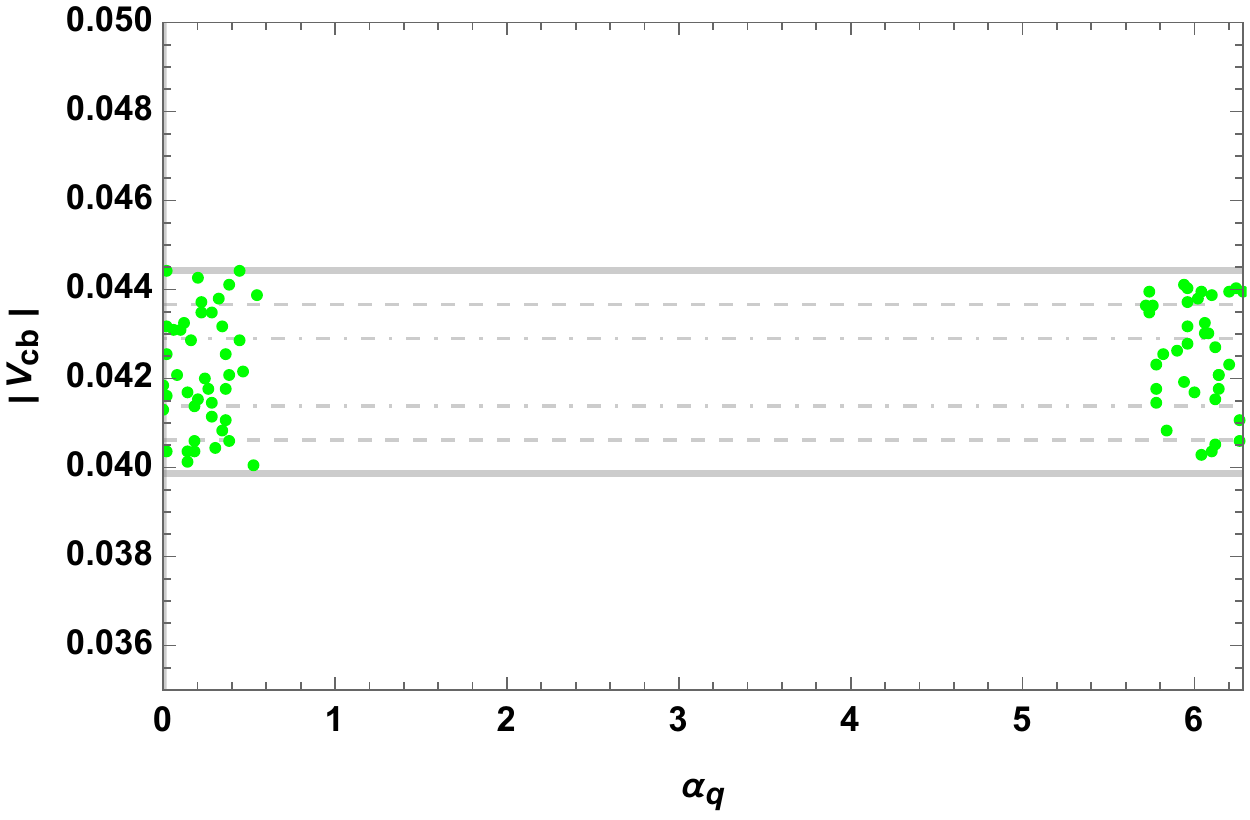}
\caption{$\vert V_{cb}\vert$ versus $\tilde{a}_{u}$, $\tilde{a}_{d}$ and $\bar{\alpha}_{q}$, respectively. The dotdashed, dashed and thick lines stand for $1~\sigma$, $2~\sigma$ and $3~\sigma$.}\label{fig1} 
\end{figure}

The fig.\ref{fig1} shows that there is region of values for the free parameters where the entry $\vert V_{cb}\vert$ is consistent with the experimental values. A remarkable fact is that for the CP phase there are two small region where $\bar{\alpha}_{q}$ is close to $0$ and $2\pi$, as one can see. The rest of the plots, $\vert V_{ui}\vert$ ($i=d,s,b$), will not show because these are redundant since these will be within the experimental regions since that we demanded it.

\begin{figure}[h!]
	\centering
	\includegraphics[scale=0.465]{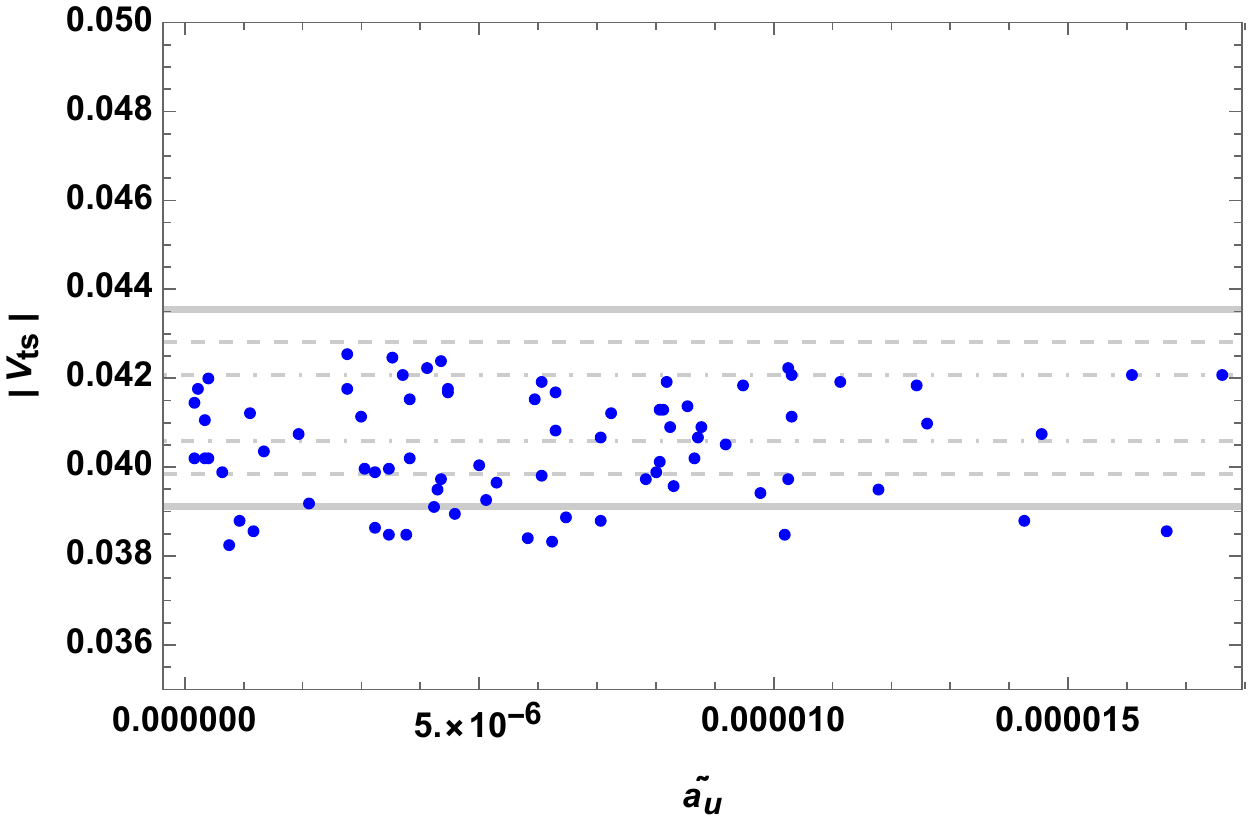}\hspace{1mm}\includegraphics[scale=0.465]{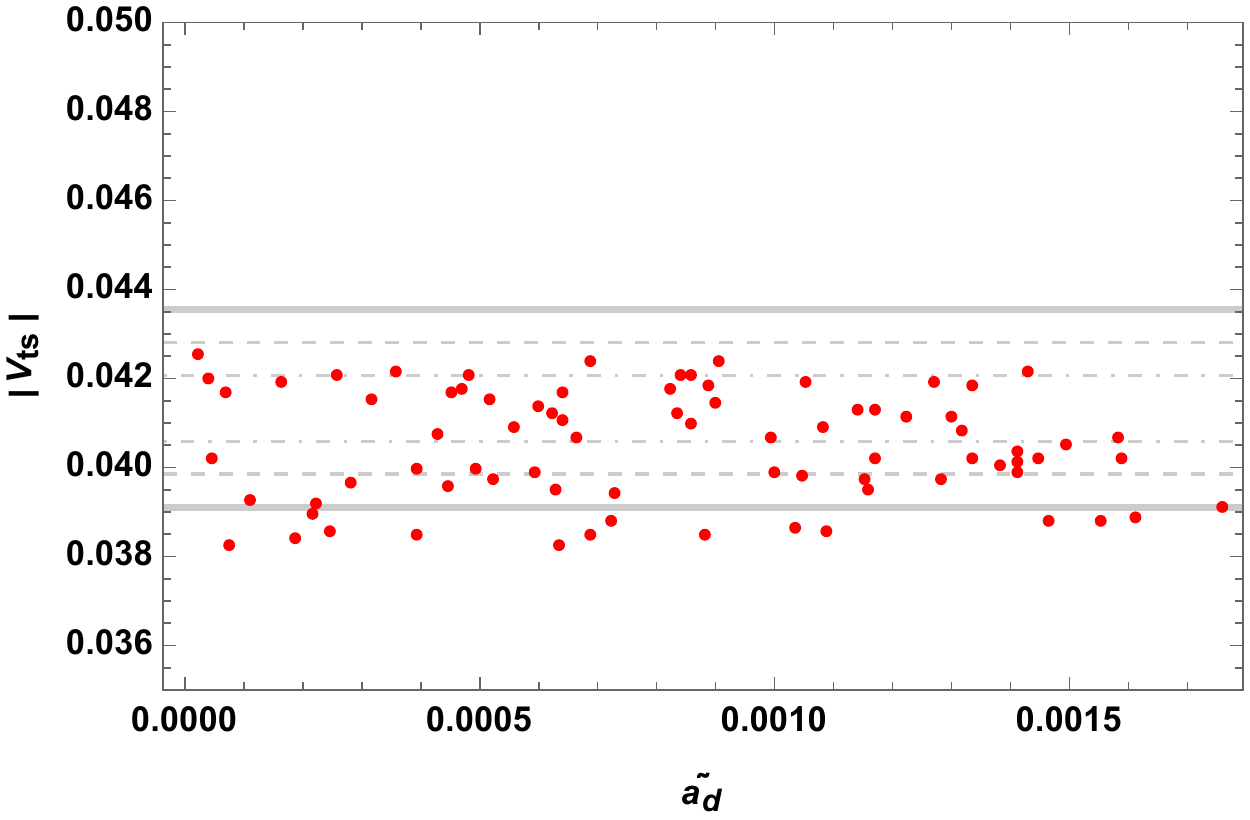}\hspace{1mm}\includegraphics[scale=0.465]{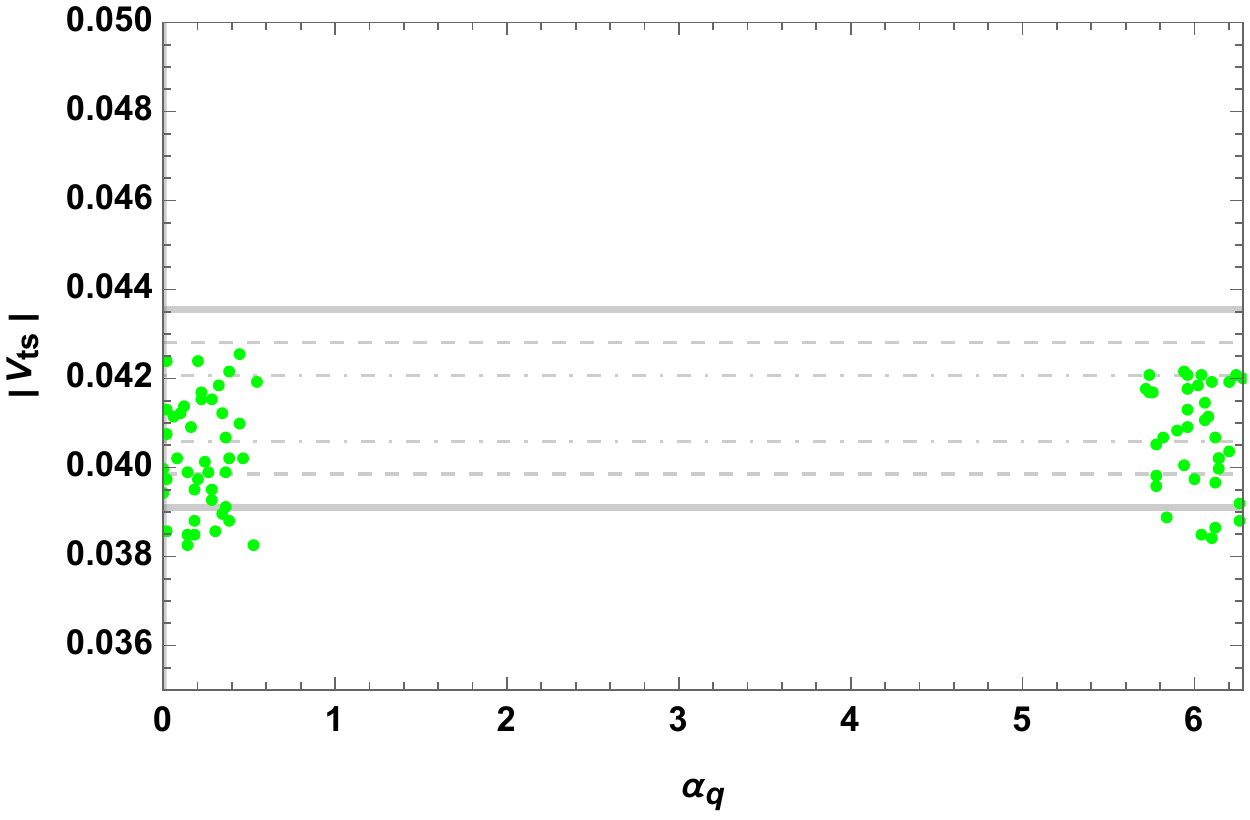}
	\caption{$\vert V_{ts}\vert$ versus $\tilde{a}_{u}$, $\tilde{a}_{d}$ and $\bar{\alpha}_{q}$, respectively. The dotdashed, dashed and thick lines stand for $1~\sigma$, $2~\sigma$ and $3~\sigma$.
}\label{fig2} 
\end{figure}

Let us include some plots that are model predictions. First, the fig. (\ref{fig2}) shows the permitted region for the observable $\vert V_{ts} \vert$ with the same allowed region for the free parameters. 

\begin{figure}[h!]
	\centering
	\includegraphics[scale=0.465]{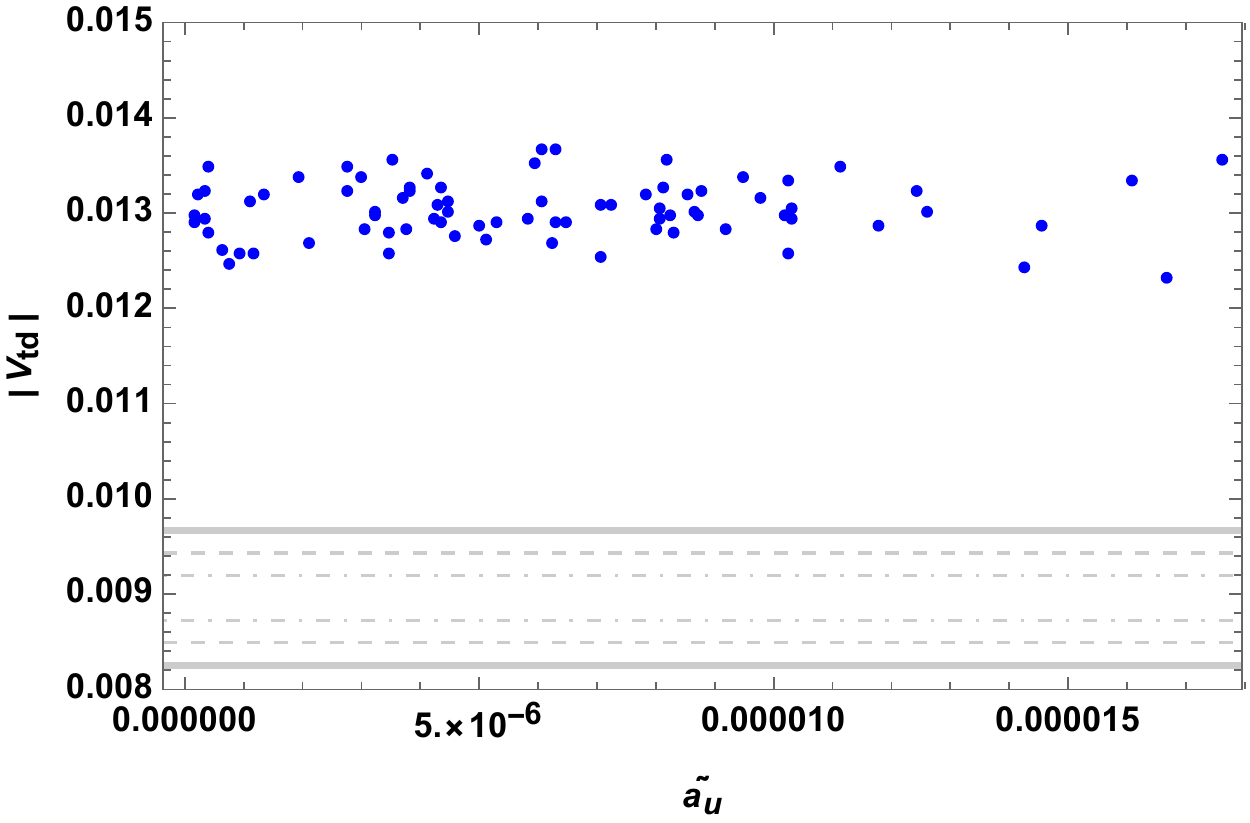}\hspace{1mm}\includegraphics[scale=0.465]{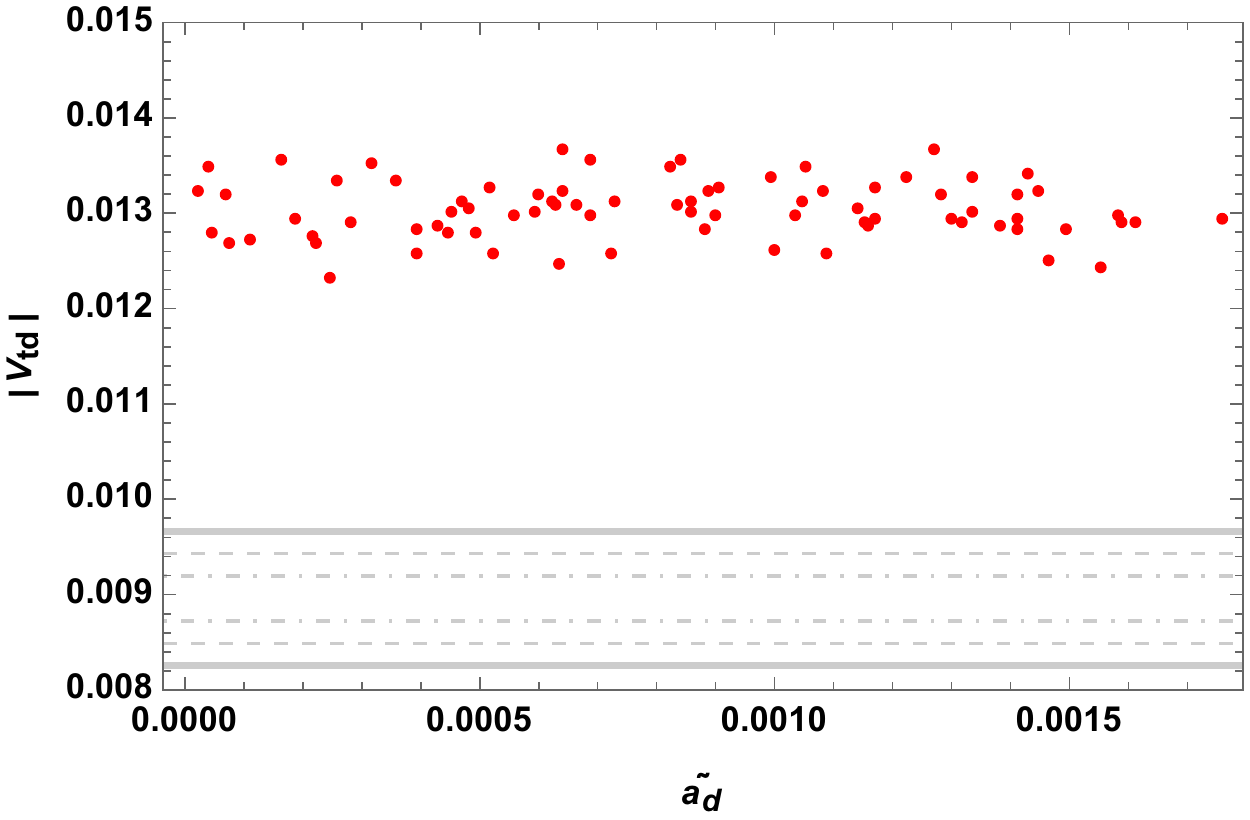}\hspace{1mm}\includegraphics[scale=0.465]{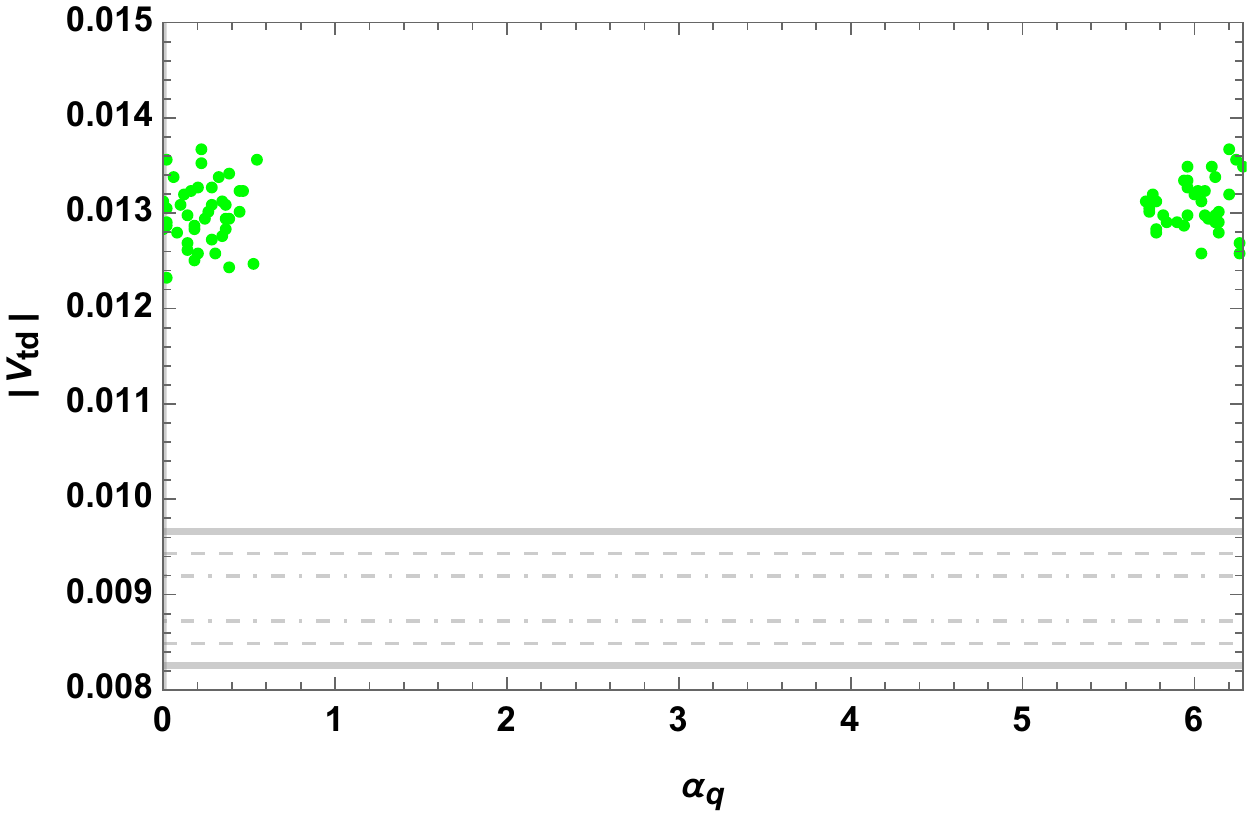}
	\caption{$\vert V_{td}\vert$ versus $\tilde{a}_{u}$, $\tilde{a}_{d}$ and $\bar{\alpha}_{q}$, respectively. The dotdashed, dashed and thick lines stand for $1~\sigma$, $2~\sigma$ and $3~\sigma$.}\label{fig3}
	\end{figure}

\begin{figure}[h!]
	\centering
	\includegraphics[scale=0.465]{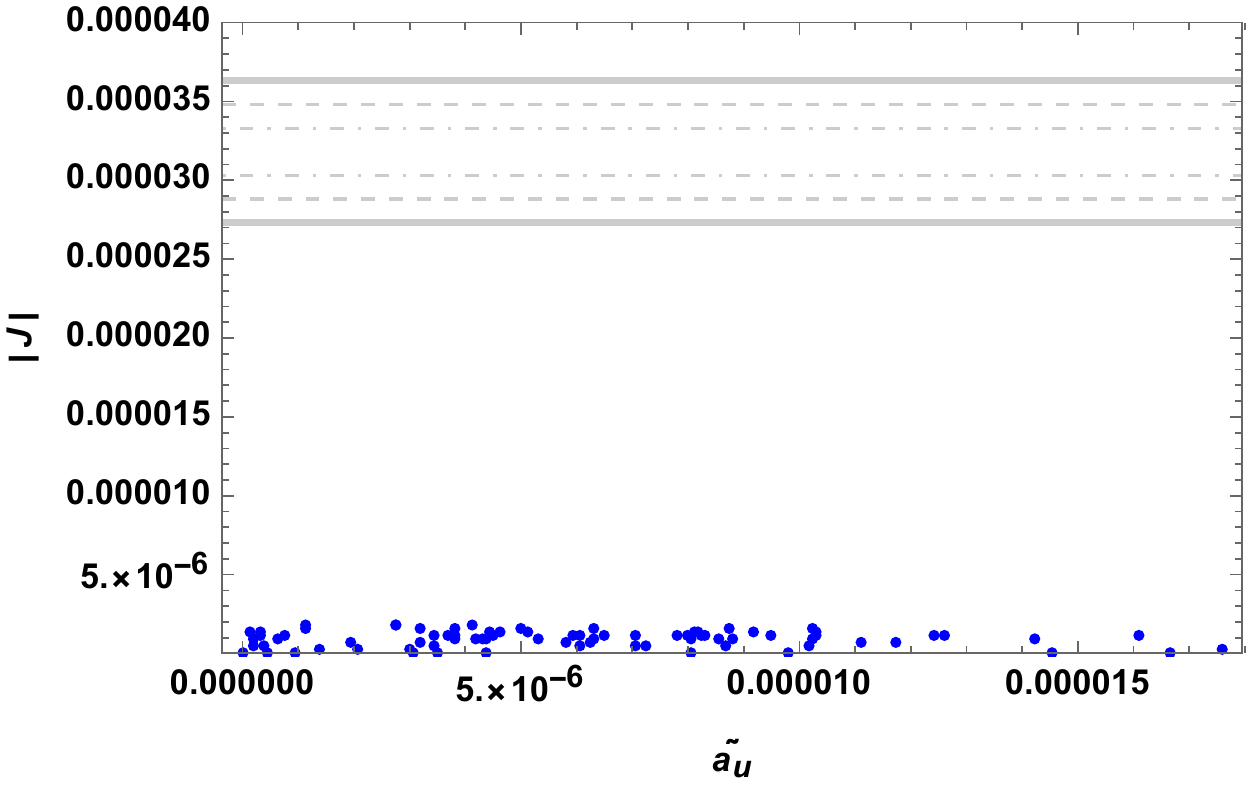}\hspace{1mm}\includegraphics[scale=0.465]{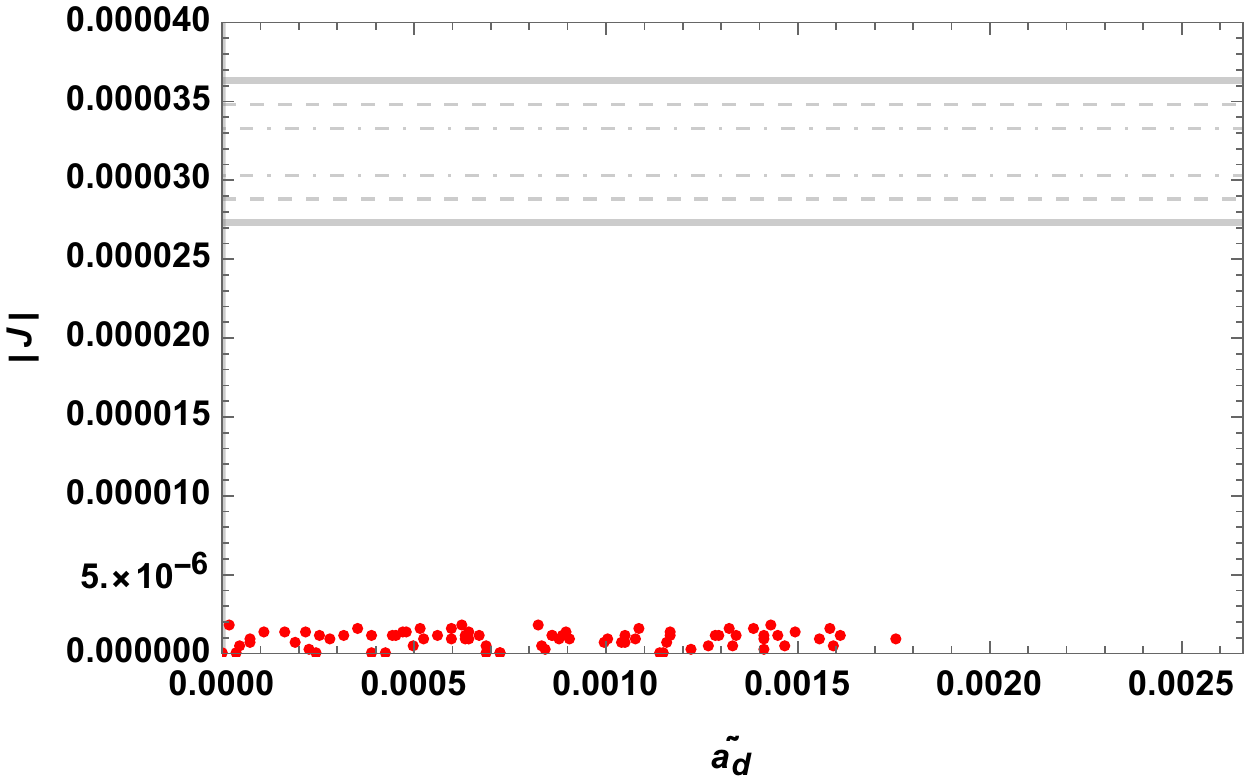}\hspace{1mm}\includegraphics[scale=0.465]{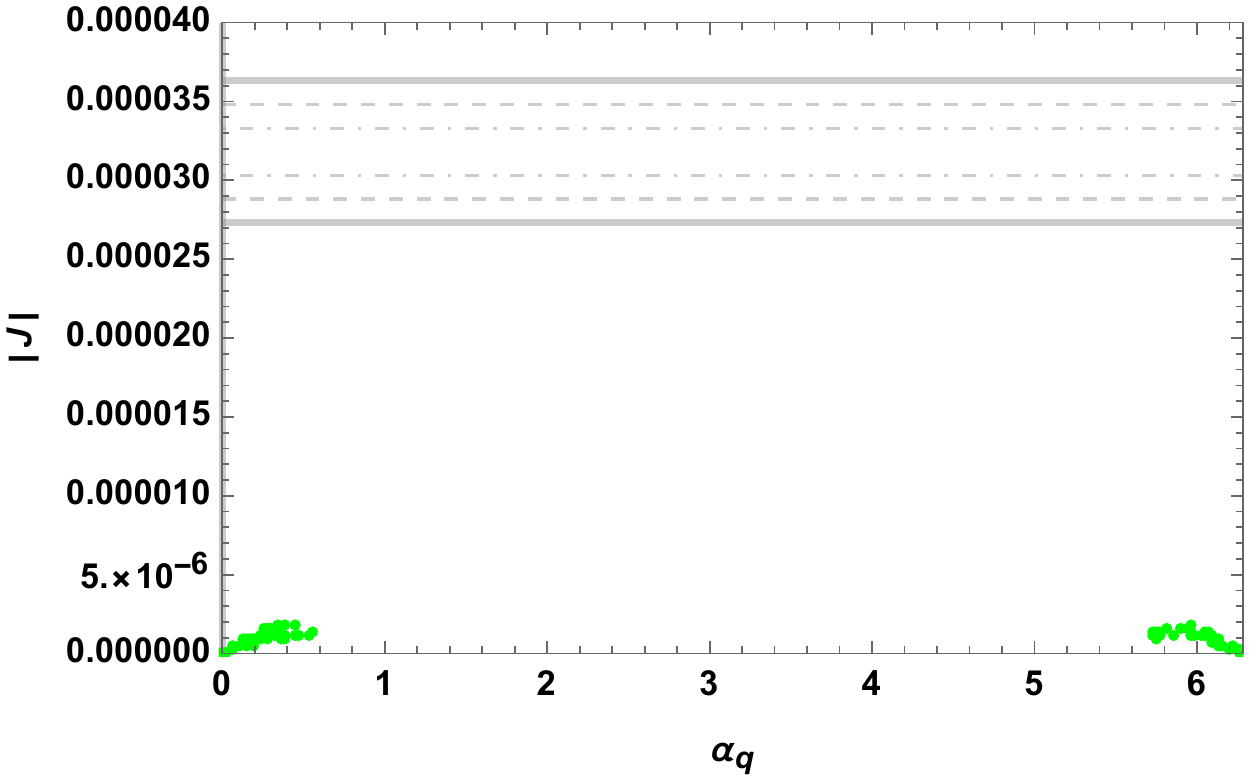}
	\caption{Jarlskog invariant versus $\tilde{a}_{u}$, $\tilde{a}_{d}$ and $\bar{\alpha}_{q}$. The dotdashed, dashed and thick lines stand for $1~\sigma$, $2~\sigma$ and $3~\sigma$.}\label{fig4} 
\end{figure}

In the figs.(\ref{fig3}-\ref{fig4}), one sees that the model predicts that entry $\vert V_{td} \vert$ and the Jarlskog invariant are respectively above and below of the experimental data. This result can be easily understood since that the mentioned observable depend strongly on the CP violation phase. More even, in the standard parametrization of the CKM mixing matrix, the corresponding CP violating phase, $\delta$, must be directly associated with the only phase of this scenario, this is, $\bar{\alpha}_{q}$. As was already mentioned, $\bar{\alpha}_{q}$ lies in two regions (close to $0$ and $2\pi$) where the Jarlskog invariant is fairly suppressed as can be seen in the figs. (\ref{fig1}-\ref{fig2}). 

\subsection{FBLM}
In this case, the CKM mixing matrix contains four free parameters namely
$y_{u}$, $y_{d}$ and two CP violating phases where the former ones satisfy the constraint $1>y_{q}>\tilde{m}_{q_{2}}>\tilde{m}_{q_{1}}$ for each sector; and the two phases are in the interval of $\left[0, 2\pi\right]$. With this in mind, then, one lets to vary the free parameters in their allowed region to accommodate the four entries of the CKM matrix. Similarity to the FLRSM scenario, we just show the plot of the entry $\vert V_{cb}\vert$ in order to not be redundant since that the other three entries are well accommodated as one can verify straight.

\begin{figure}[h!]
	\centering
	\includegraphics[scale=0.465]{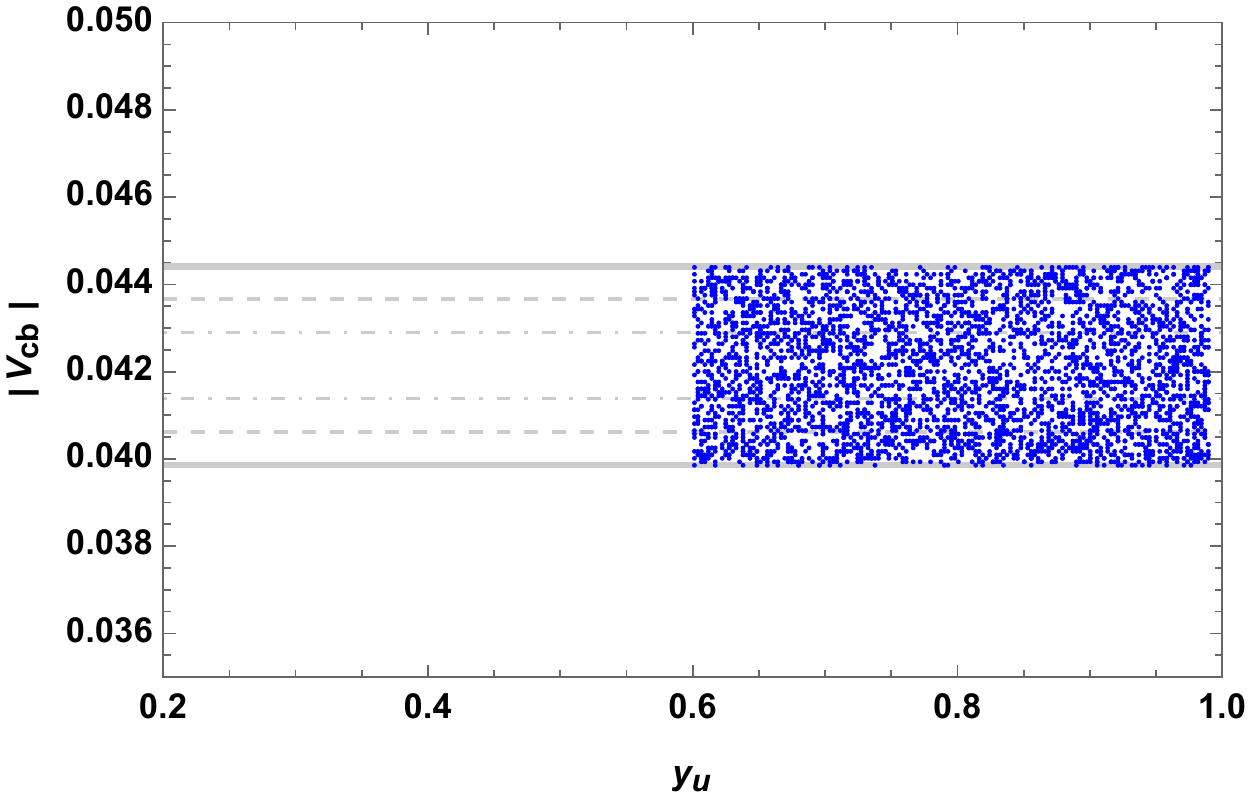}\hspace{1mm}\includegraphics[scale=0.465]{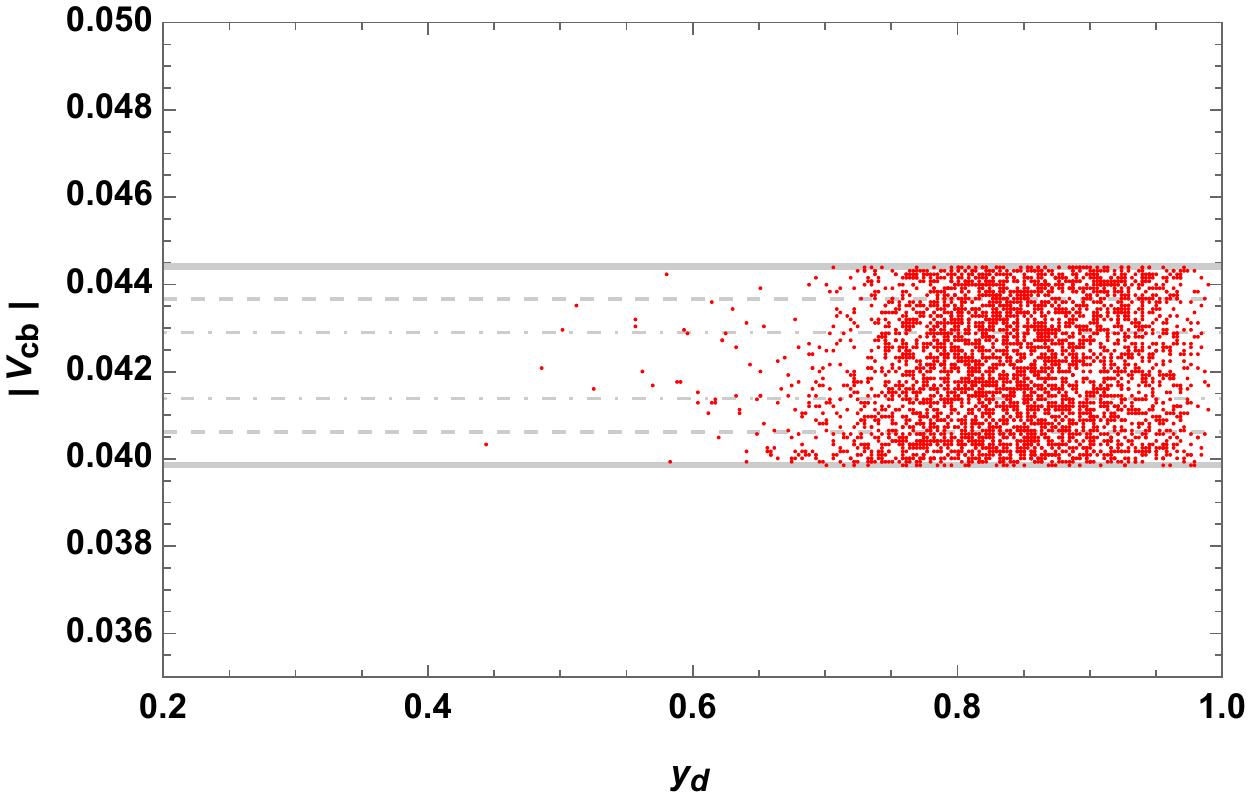} \includegraphics[scale=0.465]{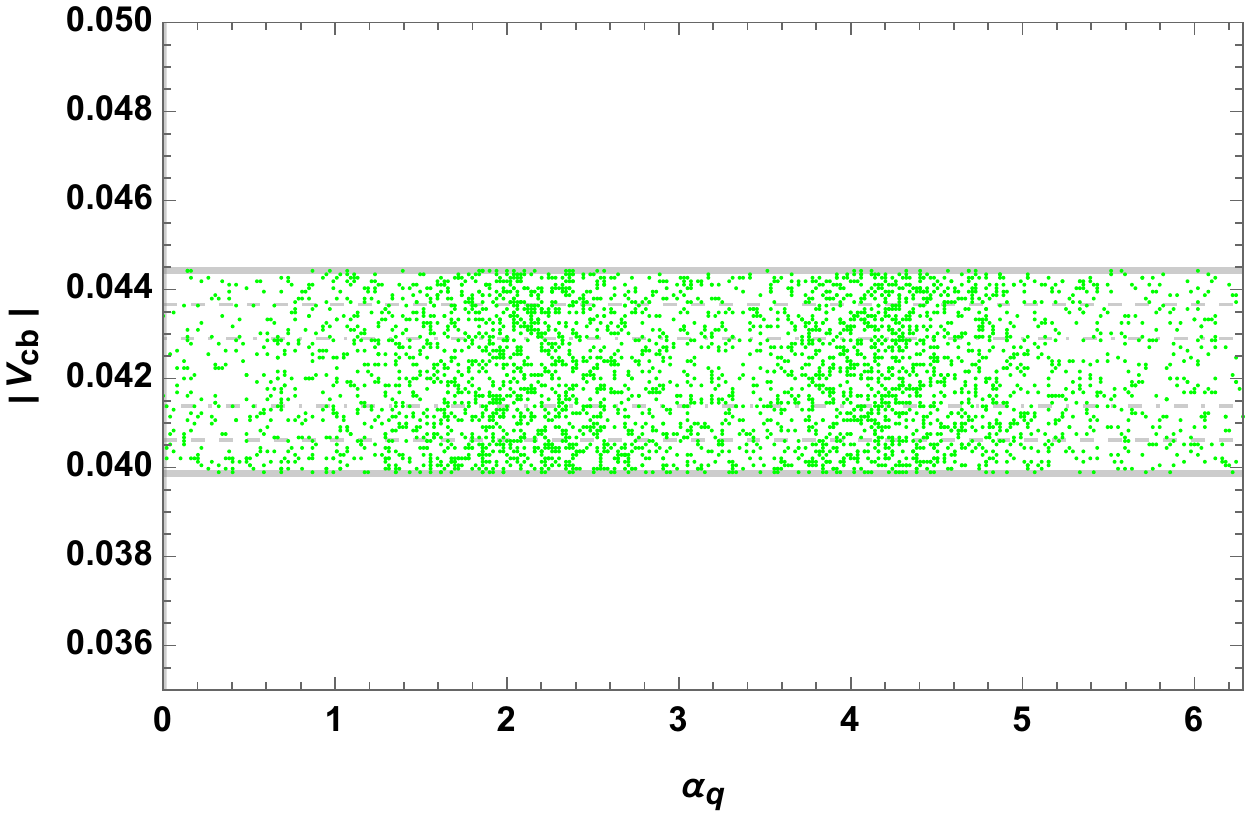}\\\includegraphics[scale=0.465]{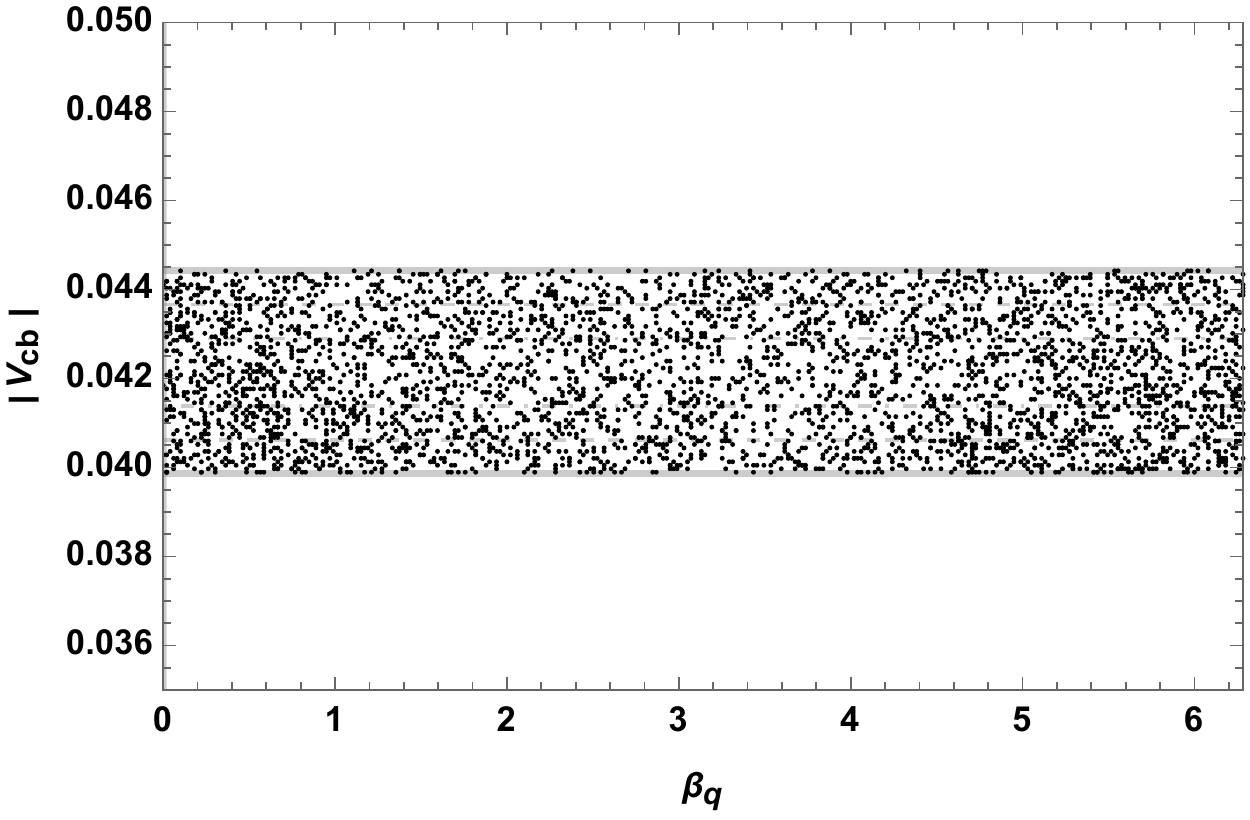}
	\caption{$\vert V_{cb}\vert$ versus the four free parameters. The dotdashed, dashed and thick lines stand for $1~\sigma$, $2~\sigma$ and $3~\sigma$.
	} \label{fig5} 
\end{figure}

As we can see, in the Fig.(\ref{fig5}), there is a region of values for the free parameters where the entry, $\vert V_{cb}\vert$, is in good agreement with experimental data. Remarkably, there are two notable regions for the CP phase, $\alpha_{q}$, where the mentioned CKM entries is accommodated. In addition, that entry does not depend strongly on the second CP violating phase, as one sees in the corresponding figure. In general, the CKM entries are non sensitives to the second CP phase, $\beta_{q}$, as we will see in the next scattering plots.

\begin{figure}[ht]
\centering
\includegraphics[scale=0.465]{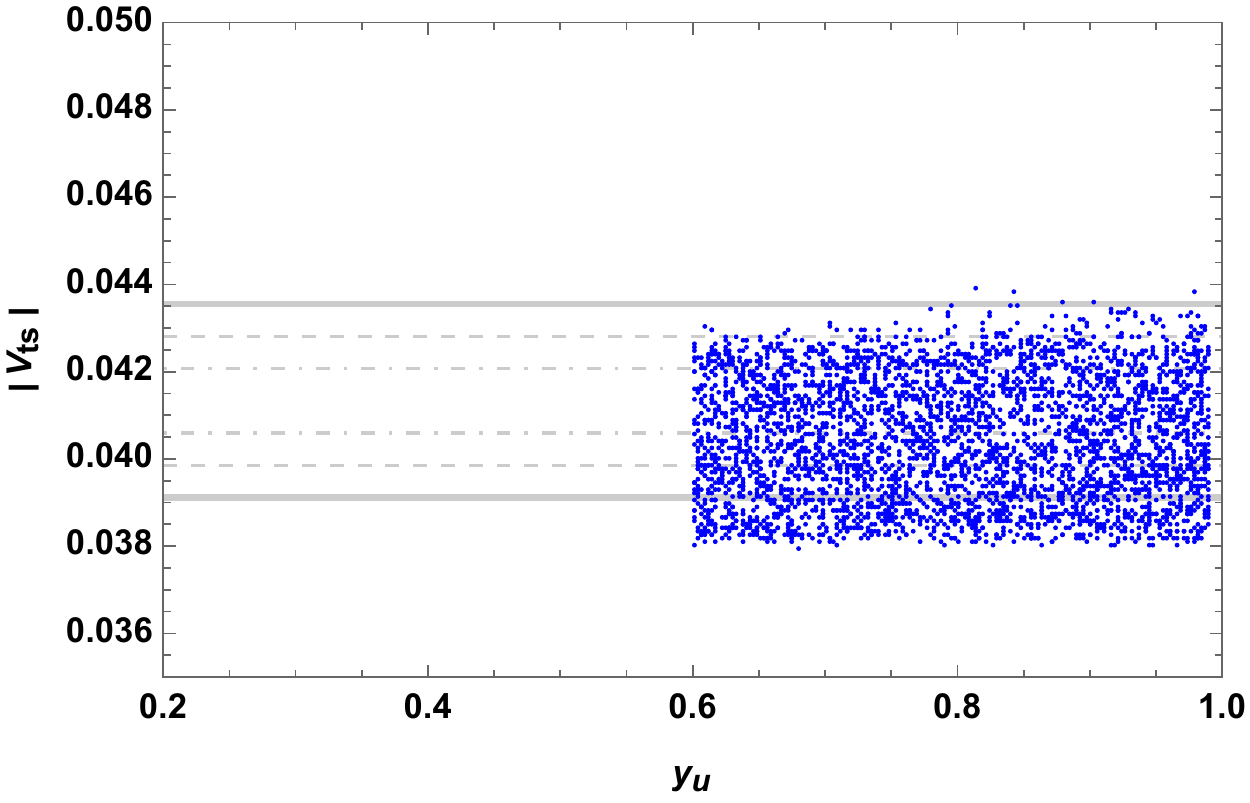}\hspace{1mm}\includegraphics[scale=0.465]{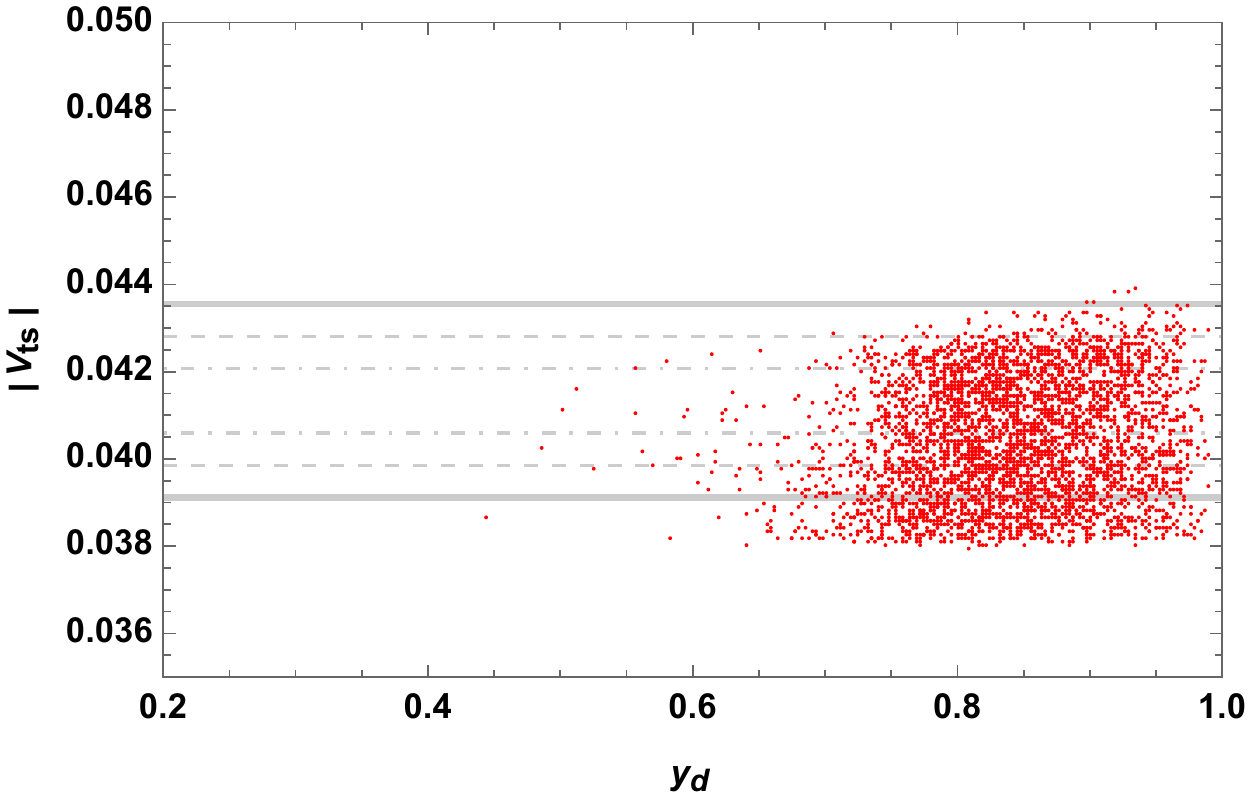} \includegraphics[scale=0.465]{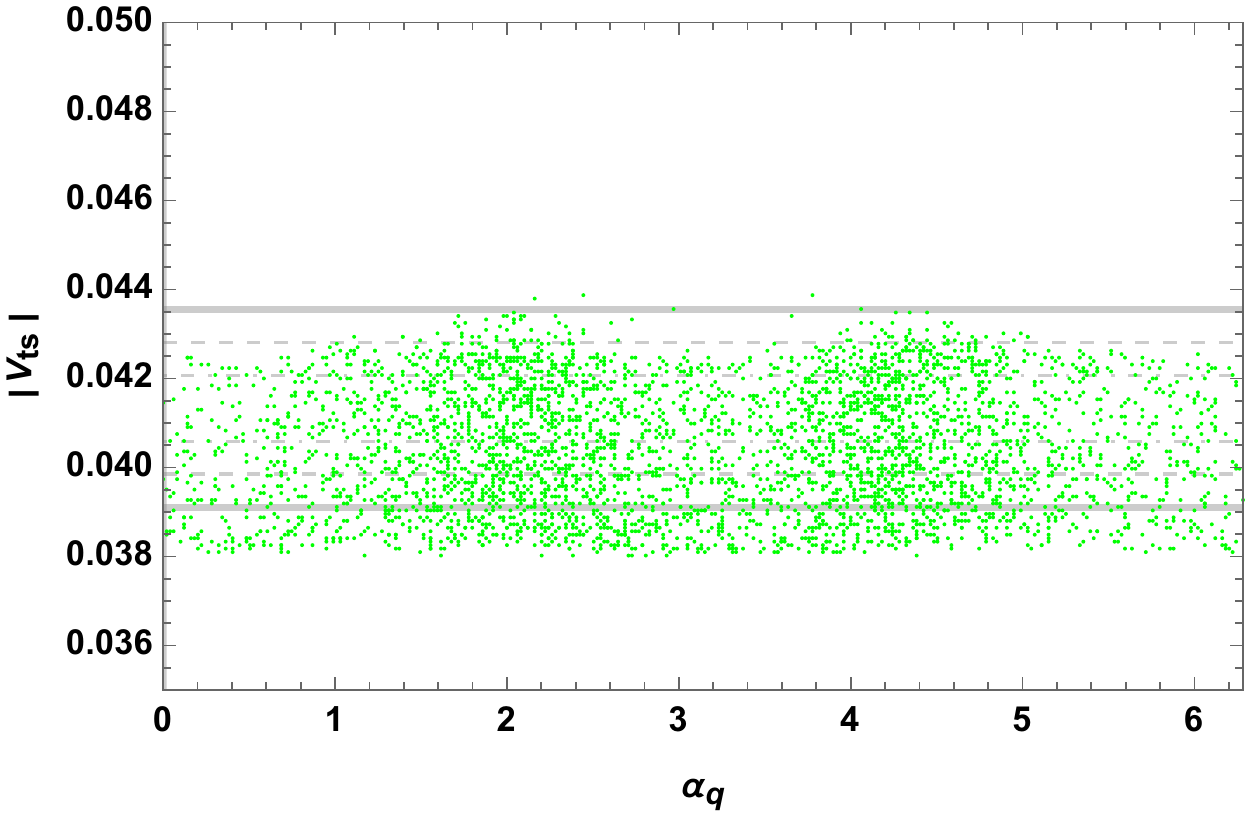}\\\includegraphics[scale=0.465]{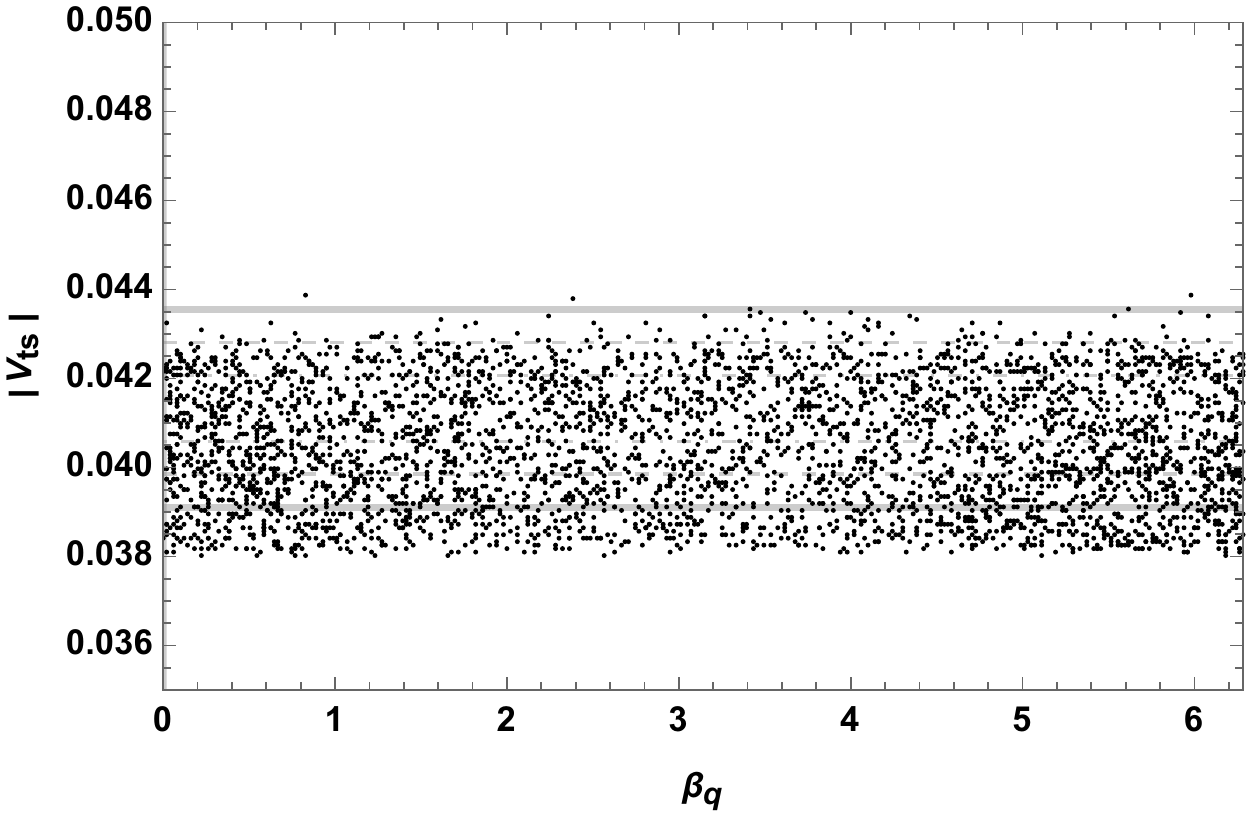}
\caption{$\vert V_{ts}\vert$ versus the four free parameters. The dotdashed, dashed and thick lines stand for $1~\sigma$, $2~\sigma$ and $3~\sigma$.
}\label{fig6} 
\end{figure}

Additionally, a model prediction, $\vert V_{ts}\vert$, is shown in the figure \ref{fig6}. As can be noticed, analogously to the entry, $\vert V_{cb}\vert$, there is a similar behavior for $\vert V_{ts}\vert$ with respect the free parameters. Let us remark that the fourth plots supports our statement on the minor role that plays the second phase in this framework.

\begin{figure}[ht]
\centering
\includegraphics[scale=0.465]{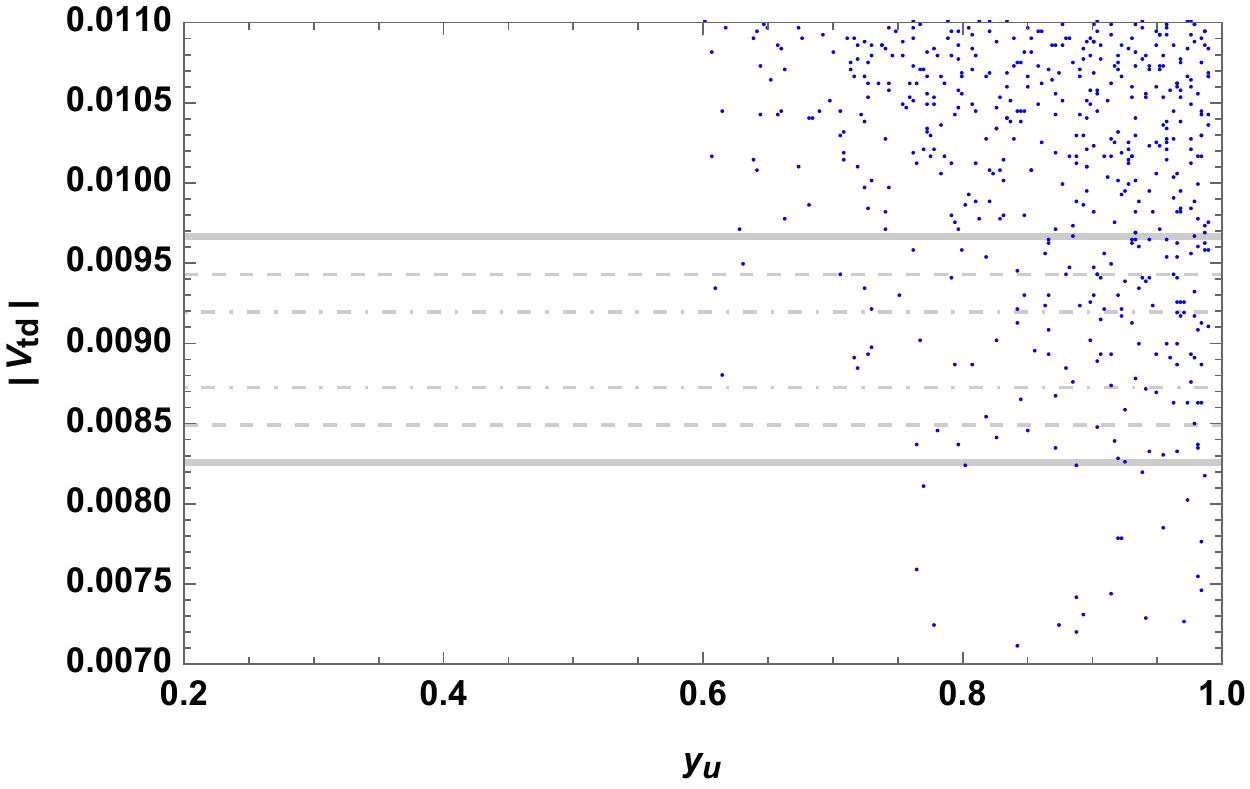}\hspace{1mm}\includegraphics[scale=0.465]{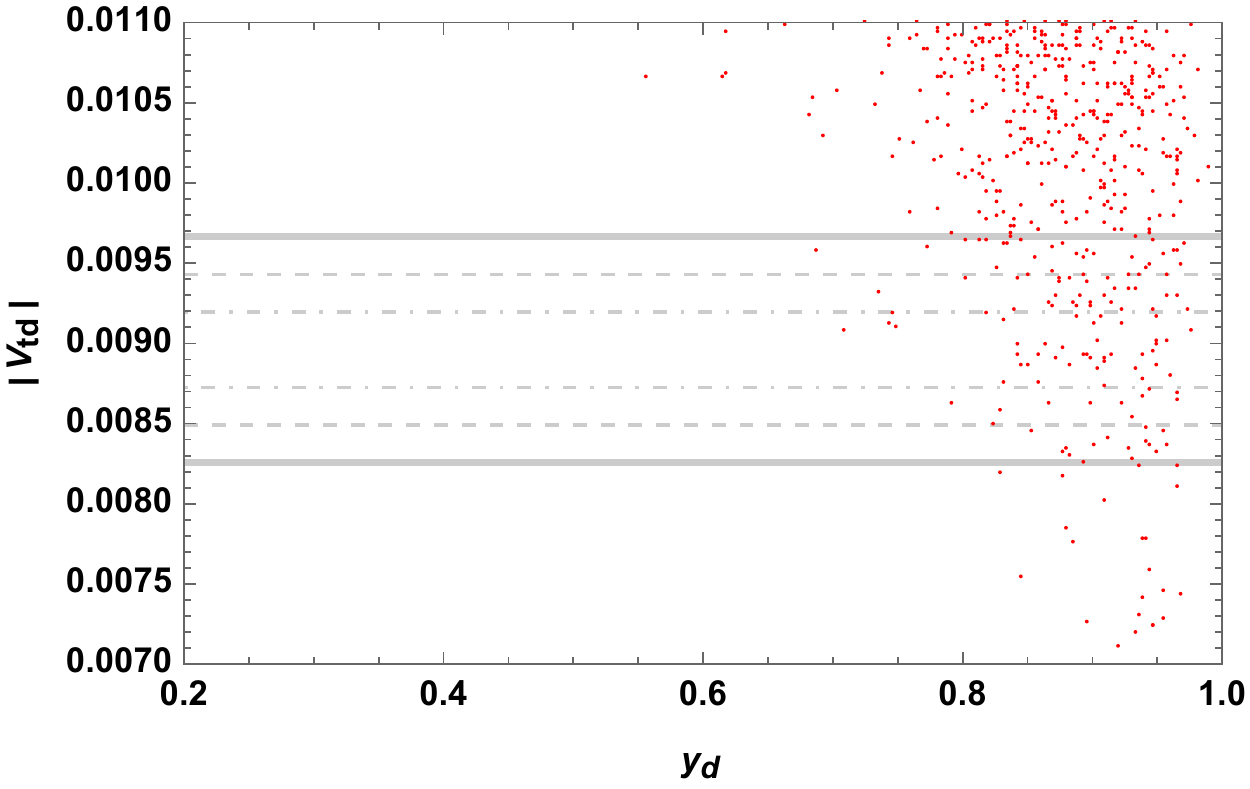} \includegraphics[scale=0.465]{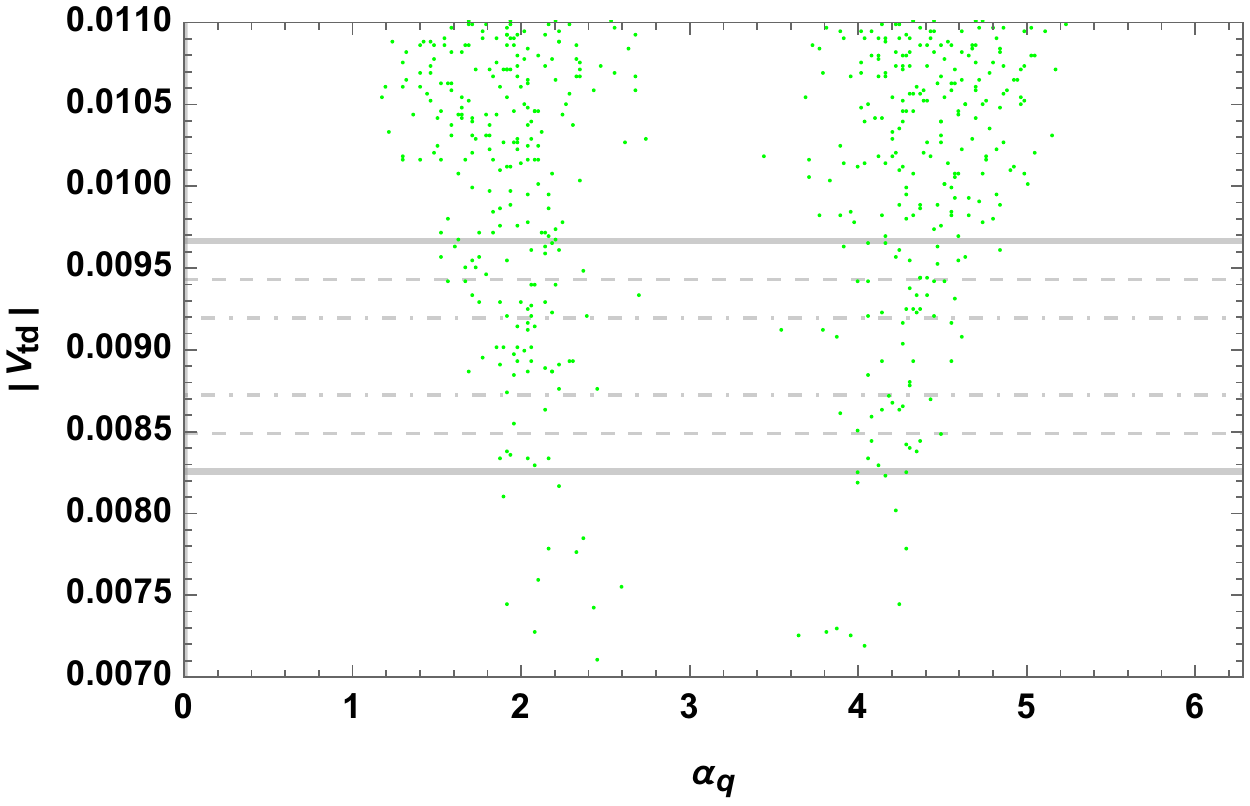}\\ \includegraphics[scale=0.465]{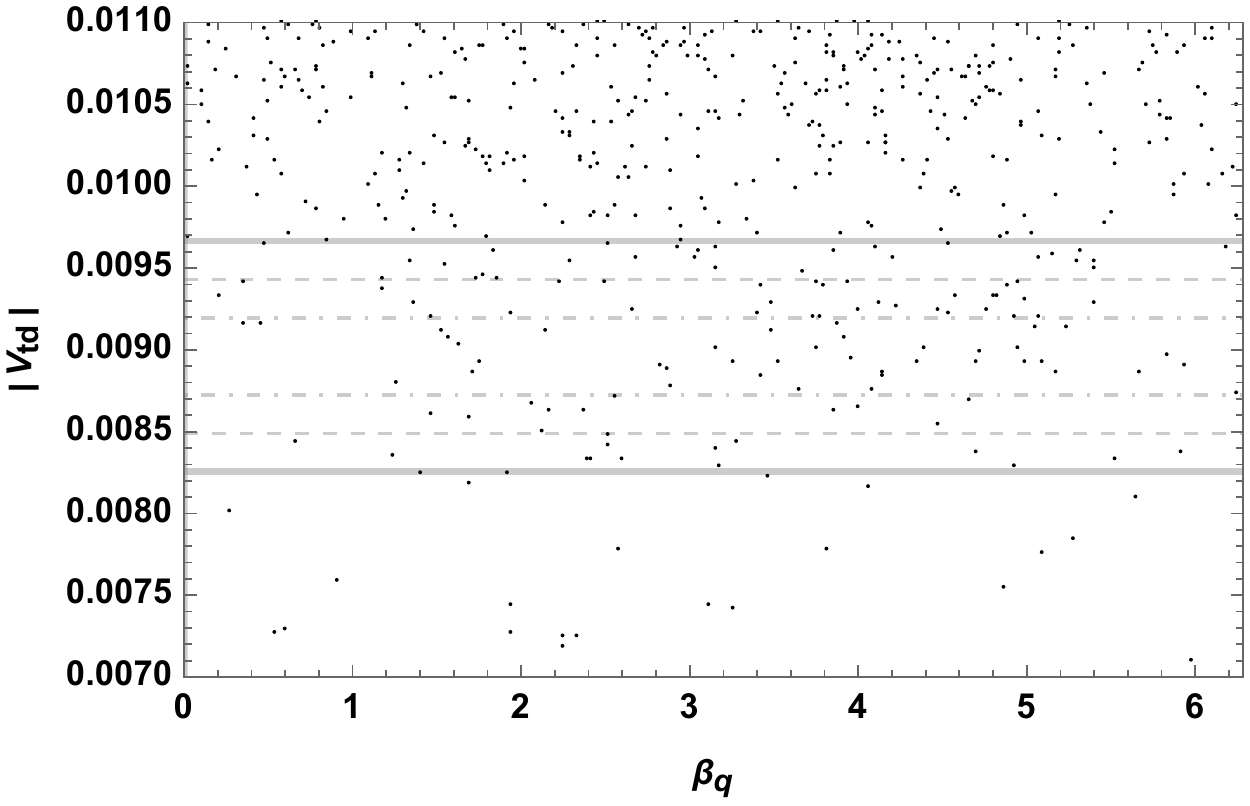}
\caption{$\vert V_{td}\vert$ versus the four free parameters. The dotdashed, dashed and thick lines stand for $1~\sigma$, $2~\sigma$ and $3~\sigma$.
}\label{fig7} 
\end{figure}

In the plot \ref{fig7}, we see the entry $\vert V_{td}\vert$ as function of the free parameters. As one can realize, this observable constraints strongly the allowed regions for the free parameters. Although there are few points, in the FBLM scenario, the mentioned CKM element is fitted quite well in comparison to the FLRSM case.

\begin{figure}[ht]
	\centering
	\includegraphics[scale=0.465]{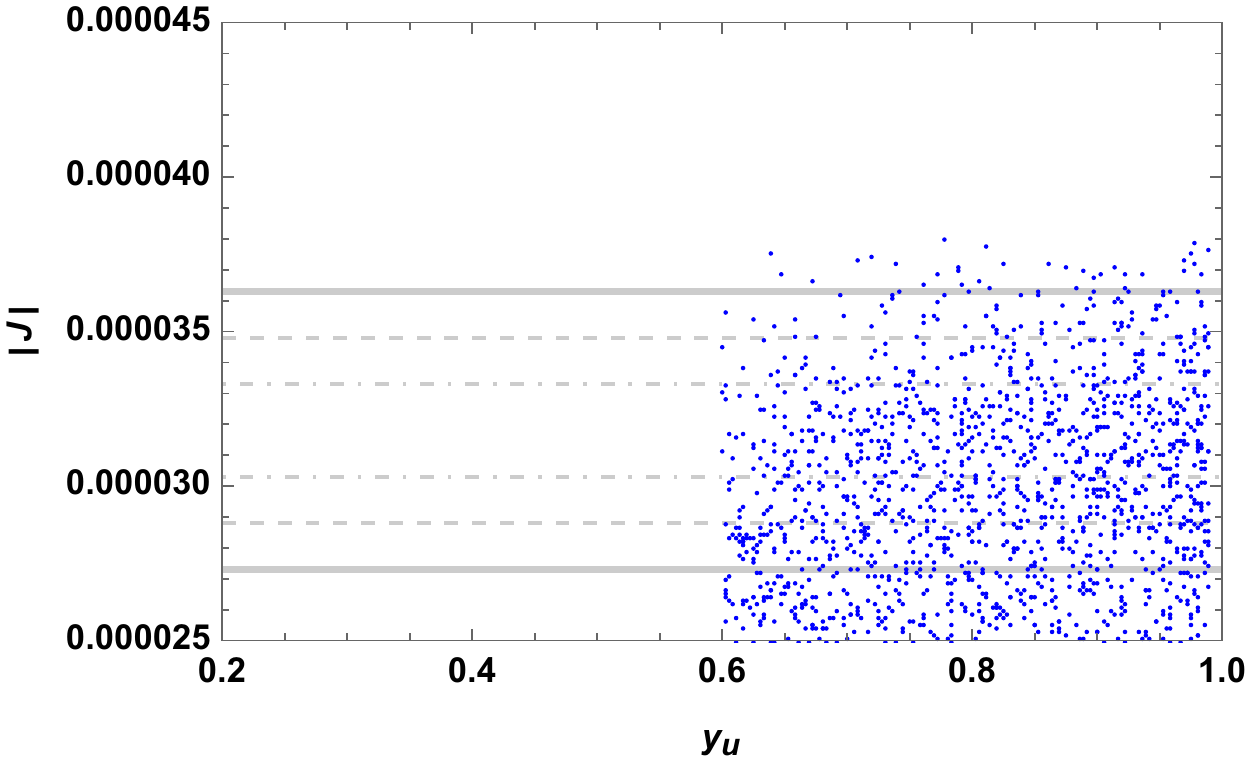}\hspace{1mm}\includegraphics[scale=0.465]{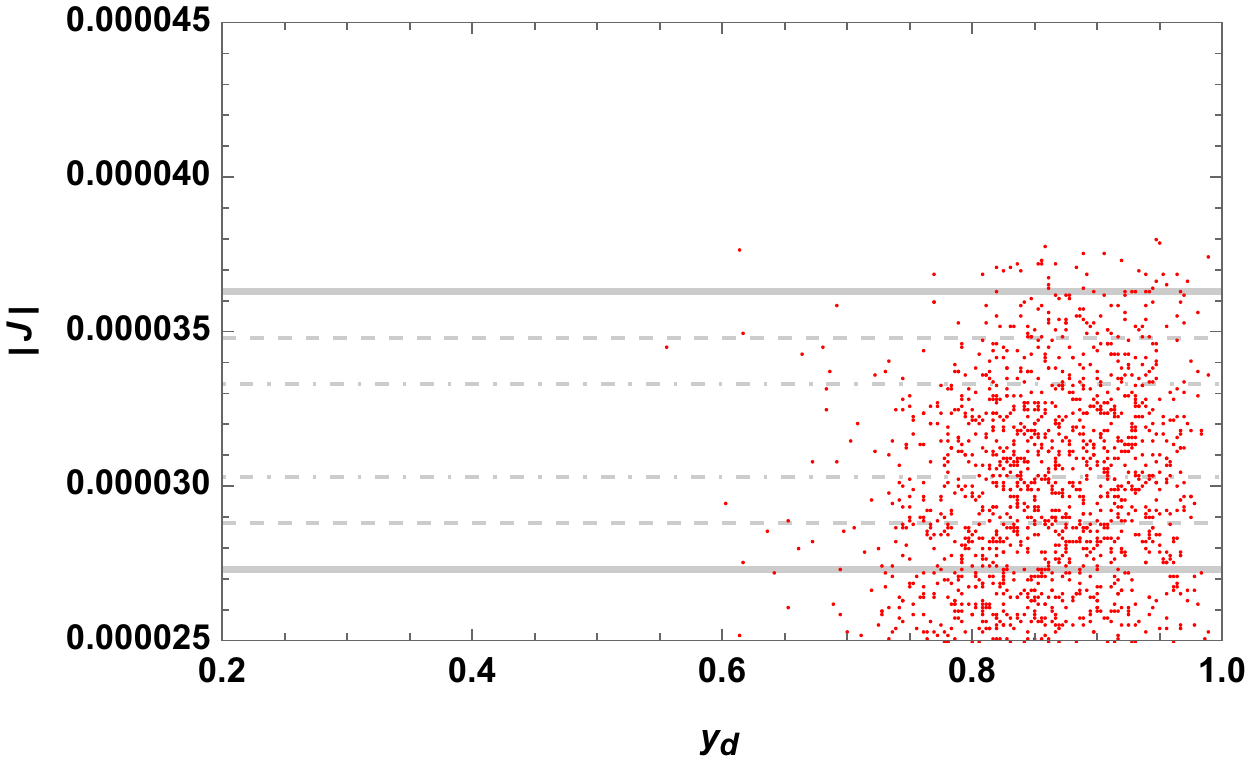} \includegraphics[scale=0.465]{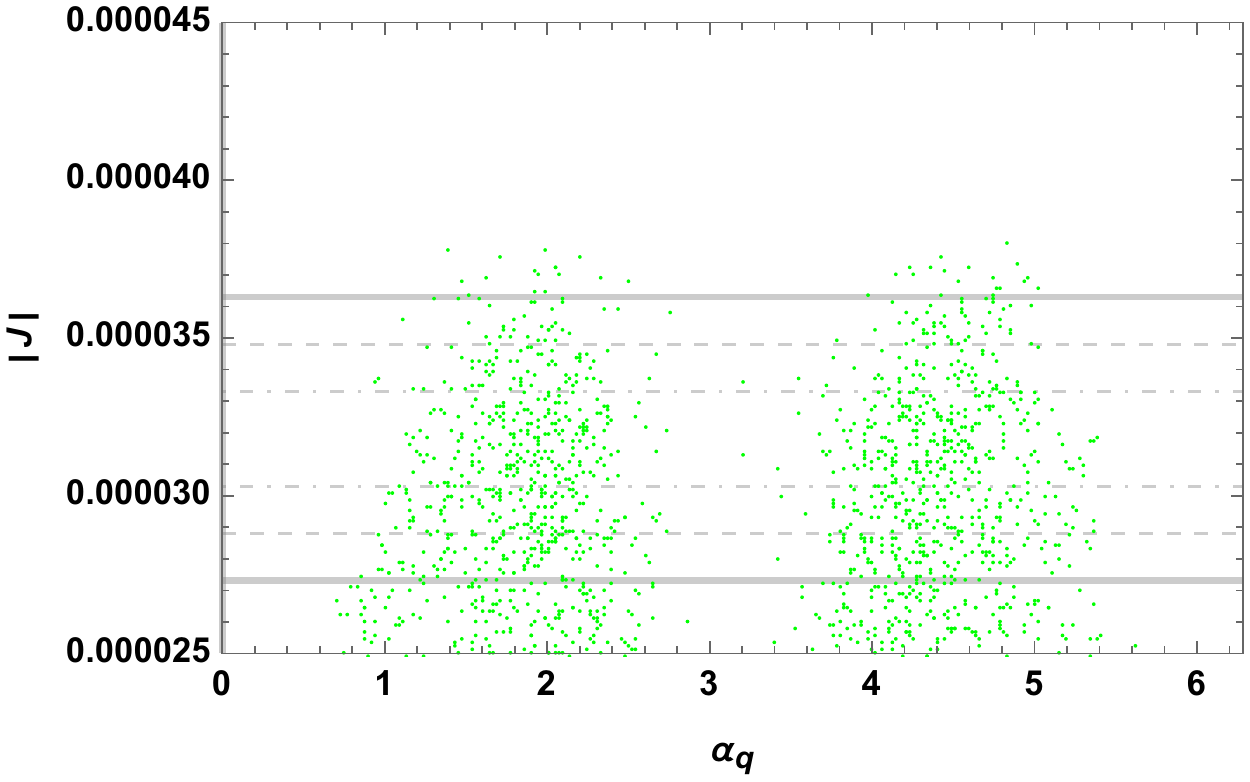}\\ \includegraphics[scale=0.465]{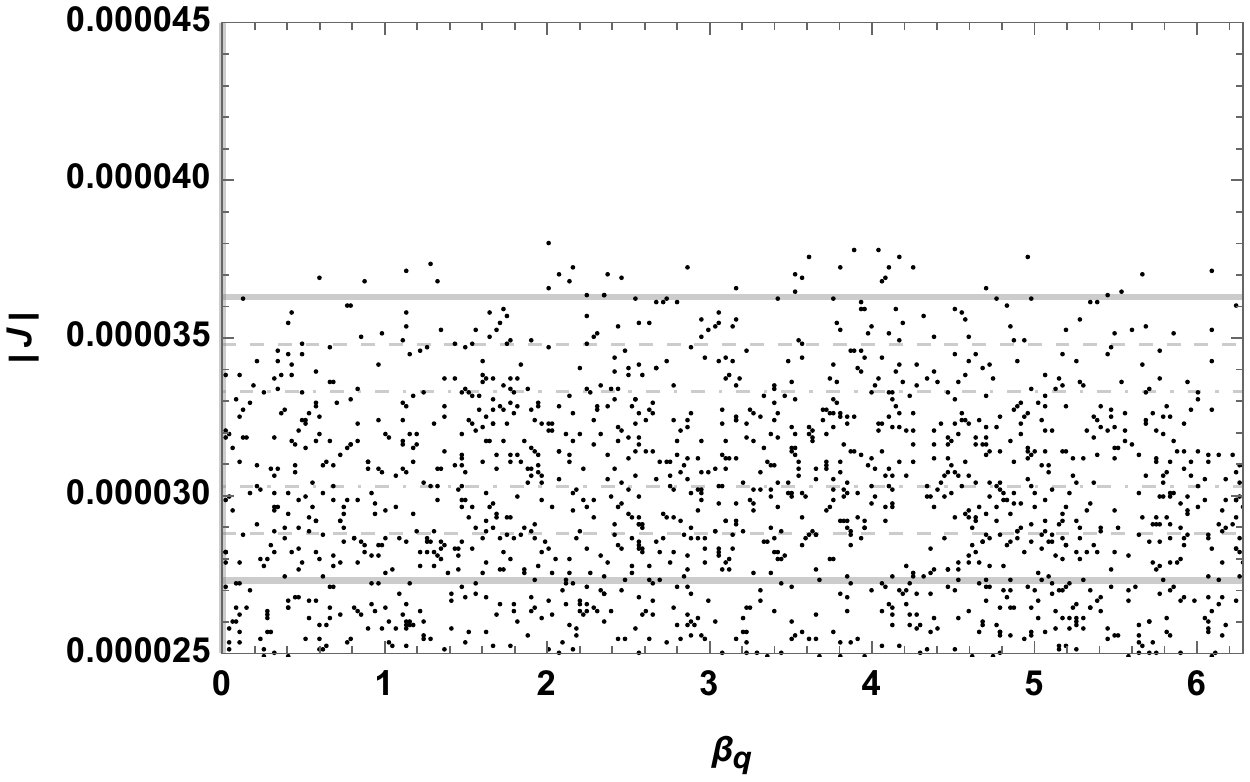}
	\caption{Jarlskog invariant versus the four free parameters. The dotdashed, dashed and thick lines stand for $1~\sigma$, $2~\sigma$ and $3~\sigma$.}\label{fig8} 
\end{figure}

At the same time, the permitted region of values for the Jarlskog invariant is displayed in the figure (\ref{fig8}). In this case, the CP violating phase, $\delta$, must be a combination of the two phase $\alpha_{q}$ and $\beta_{q}$ which are the only source of CP violation in the FBLM scenario. Remarkably, the predicted region of values for the Jarlskog invariant is compatible with the experimental results. 

Before finishing this section,
let us make some general comments on the FLRSM and FBLM frameworks. In the former one, the CP violating phase is not sufficient to fit simultaneously the
$\vert V_{td}\vert$ and the Jarlskog invariant.
In the latter case, the CP phase, $\alpha_{q}$, plays an important role in the CKM matrix elements in comparison to the second phase, $\beta_{q}$, as one can see in the above scattering plots. In fact, the CKM matrix is more sensible to the former phase than the second one. 

\section{Conclusions}

We have made a brief comparison of the quark sector that comes from two gauge models dress with the $\mathbf{S}_{3}$ flavor symmetry. In spite of the fact that those models were studied before, the presented simple scenarios have not been released neither compared in the literature.

As was shown, two simple cases, one for each model, were studied and compared taking into account their predictions on the CKM mixing matrix. Albeit the FLRSM scenario has fewer free parameters, it is not able to fit the CKM mixing matrix since that the only CP violating phase is not sufficient to accommodate the observables that depend strongly of that parameter. On the other hand, there are more free parameters in the FBLM case, however, a strong assumption on the Yukawa couplings was done to reduce the unknown parameters. In this benchmark, the NNI textures drive the quark mixings and, as we noticed,  there are allowed regions for the free parameters where CKM mixing matrix is in good agreement with experimental data. In addition, the predicted region for the Jarlskog invariant lies withing the experimental limits. 

According the obtained results, the NNI textures work better than the modified Fritzsch under the assumptions carried out.

\section*{Acknowledgements}
Garc\'ia-Aguilar appreciates the facilities given by the IPN through the SIP project number 20195636. JCGI thanks to Instituto Polit\'enico Nacional (IPN) for being benefited with the Project SIP 20196559. This work was partially supported by the Mexican grants 237004, PAPIIT IN111518 and Conacyt-32059.

\appendix{}

\appendix
\section{$\mathbf{S}_{3}$ flavour symmetry}

The non-Abelian group ${\bf S}_{3}$ is the permutation group of three objects \cite{Ishimori:2010au} and this has three irreducible representations: two 1-dimensional, ${\bf 1}_{S}$ and ${\bf 1}_{A}$, and one 2-dimensional representation, ${\bf 2}$. The multiplication rules among them are: ${\bf 1}_{S}\otimes {\bf 1}_{S}={\bf 1}_{S}$, ${\bf 1}_{S}\otimes {\bf 1}_{A}={\bf 1}_{A}$ and ${\bf 1}_{A}\otimes {\bf 1}_{A}={\bf 1}_{S}$; ${\bf 1}_{S}\otimes {\bf 2}={\bf 2}$, ${\bf 1}_{A}\otimes {\bf 2}={\bf 2}$ and the last one

\begin{eqnarray}\label{rules}
\begin{pmatrix}
a_{1} \\ 
a_{2}
\end{pmatrix}_{{\bf 2}}
\otimes
\begin{pmatrix}
b_{1} \\ 
b_{2}
\end{pmatrix}_{{\bf 2}}=
\left(a_{1}b_{1}+a_{2}b_{2}\right)_{{\bf 1}_{S}} \oplus  \left(a_{1}b_{2}-a_{2}b_{1}\right)_{{\bf 1}_{A}} \oplus	
\begin{pmatrix}
a_{1}b_{2}+a_{2}b_{1} \\ 
a_{1}b_{1}-a_{2}b_{2}
\end{pmatrix}_{{\bf 2}}. 
\end{eqnarray}

\bibliographystyle{bib_style_T1}
\bibliography{references.bib}

\end{document}